\documentclass[a4paper,11pt]{article}
\usepackage{jheppub} 
\usepackage{lineno}
\usepackage{nicefrac}
\usepackage{etoolbox}
\usepackage{placeins}
\usepackage{appendix}

\usepackage[normalem]{ulem}

\makeatletter
\patchcmd{\thebibliography}{\advance\leftmargin\labelsep}{\leftmargin\z@\itemindent\z@}{}{}
\patchcmd{\@biblabel}{\hss}{\hfill}{}{}
\makeatother

\title{\boldmath Gravitational Waves From an Entangled $SO(3)$ Cosmic String Network}

\author{Mu-Chun Chen,}
\author{Cameron Moffett-Smith}
\affiliation{Department of Physics and Astronomy, 
University of California, Irvine, 
California, U.S.A}

\emailAdd{muchunc@uci.edu}
\emailAdd{cmmoffet@uci.edu}

\begin{document}
\begin{flushright}
UCI-TR-2026-19
\end{flushright}
\abstract{Discrete non-Abelian symmetries such as $A_4$ are widely used in models of lepton flavor to explain neutrino mass hierarchy and their large mixing. If such a symmetry arises from the spontaneous breaking of a continuous group, cosmic strings can form. We study the breaking $\mathrm{SO}(3)\to A_4$, for which the strings are classified by the non-Abelian group $\pi_1(\rm{SO}(3)/A_4)\cong T'$, the binary tetrahedral group. Because the string fluxes do not commute, colliding strings may reconnect, pass through each other, or become joined by a bridge whose flux can belong to a different string network. We derive the probabilities of these outcomes directly from the group structure; in network models they are usually treated as free parameters. Including them in a multi-network velocity-dependent one-scale model, we find that the correlation length of the networks decreases and their energy density increases relative to Abelian $\mathrm{U}(1)$ strings of the same tension. The resulting stochastic gravitational-wave background is enhanced by up to nearly three orders of magnitude, showing that the algebraic structure of $\pi_1$ leaves a direct imprint on the gravitational-wave spectrum.}
\maketitle

\section{Introduction}
The Standard Model (SM) is one of the most successful theories of nature describing the strong nuclear force, the weak nuclear force, and the electromagnetic force. However there are questions unanswered in the SM such as the hierarchical masses of the quarks and  leptons, the origin of neutrino masses, and the form of the Pontecorvo-Maki-Nakagawa-Sakata (PMNS) matrix being non-diagonal for the leptons whereas the Cabibbo-Kobayashi-Maskawa (CKM) matrix is nearly diagonal for the quarks.  These together are referred to as the Flavor Puzzle. There are many attempts to resolve the Flavor Puzzle by extending the SM to include an extra flavor symmetry $G_{SM}\times G_F$ where $G_{SM}$ is the SM gauge group $SU(3)_c\times SU(2)_L \times U(1)_Y$, and $G_F$ is the extra flavor symmetry used to explain aspects of the Flavor Puzzle. The flavor symmetry, $G_F$ can, for example, be an extra $U(1)$ to explain mass hierarchies~\cite{Froggatt:1978nt}, or non-Abelian discrete symmetries and their modular variants such as $A_4,\,A_5,\, \text{and}    \,S_4$~\cite{Ma:2001dn,Feruglio:2017spp,Almumin:2022rml,Feruglio:2008ht,Kobayashi:2023zzc} to explain realistic neutrino mixing. A natural question that arises is the origin of such symmetries. The study of the breaking of SO(3) to non-Abelian subgroups and the subsequent phenomenology was discussed in~\cite{King:2018fke}. In this work, we extend such a study by looking at cosmological signatures of symmetry breaking.
\par
In order to obtain realistic leptonic mixing, one must introduce extra scalars, called flavons, which must gain vacuum expectation values (vevs) to spontaneously break the flavor symmetry as it has not been observed at energy scales that have been probed so far. In the spontaneous breaking of a group G to its subgroup H, we define the resultant vacuum manifold $\mathcal{M}=G/H$, which in general is a set with interesting topological properties. The homotopy groups, $\pi_n$, are mappings from an n-dimensional sphere $S^n$ into the vacuum manifold $\mathcal{M}$~\cite{Vilenkin:2000jqa}. These homotopy groups classify what topological defects one can get from spontaneous symmetry breaking. If $\pi_0(\mathcal{M)\neq I}$ domain walls are formed; if $\pi_1(\mathcal{M)\neq I}$ cosmic strings will form; if $\pi_2(\mathcal{M)\neq I}$ monopoles form; and if $\pi_3(\mathcal{M)\neq I}$ then textures form. In this work, we will only look at cosmic strings. The dynamics of topological defects create gravitational waves that have been shown to be strongly correlated with neutrino physics in the context of discrete, non-Abelian discrete, and Abelian symmetries~\cite{Ghoshal:2026hev,Ghoshal:2025gci, Jana:2025vyb,Jueid:2023cgp,Chigusa:2018hhl,Gelmini:2020bqg,Fu:2024jhu, Fu:2025qhf,Barreto:2026igt, Dror:2019syi,Buchmuller:2019gfy,Priya:2026vpo,Puyam:2026eyi}.
\par

The gravitational wave spectrum of topological defects has been studied extensively~\cite{Zeldovich:1974uw,Vilenkin:1981zs,Barman:2022yos,Saikawa:2017hiv,Paul:2020wbz,Notari:2025kqq,Antusch:2026msw,Dunsky:2021tih,Hua:2026mgn}, providing a link between gravitational physics and particle physics. Notably, cosmic strings must show up in the symmetry breaking schemes introduced above as $\pi_1(SO(3)/H) \neq I$ for $H= \rm A_4$. As will be shown, there are interesting consequences of $SO(3)$ breaking to a discrete subgroup that we will discuss in the following paper. 

\par
\begin{figure}[h]
    \centering
    \includegraphics[width=\textwidth]{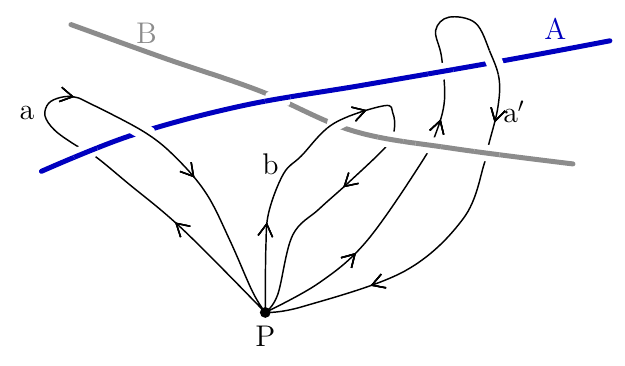}
    \caption{The paths a and  a$^{\prime}$ are not homotopically equivalent as one path cannot be continuously deformed into the other due to the presence of defect B. Figure inspired by~\cite{Avgoustidis:2007aa}.}
    \label{fig:stringsAB}
\end{figure}

 An important aspect of $SO(3)/H$ where $H=\rm A_4$, is that it is non-Abelian. This means that the definition for the flux of a cosmic string in the Abelian case
\begin{equation}
f=P\,\text{exp}(i\oint _{\gamma}A_{\mu}dx^{\mu})
\end{equation} 
is no longer gauge invariant since for $h \in H$, the flux gets transformed as $f\rightarrow hfh^{-1}$. If we have two paths, a and $\text{a}'$ enclosing the same defect with a common endpoint that cannot be continuously deformed into one another, say by the presence of another defect with path b traversing it with flux $\beta$, then they in general do not have the same flux (see Fig. \ref{fig:stringsAB}). We label such fluxes $\alpha$ and $\alpha'$. However, the two fluxes are conjugate to one another, through the intermediate defect, 
\begin{align}
    \alpha'=\beta \alpha \beta^{-1} \, ,
\end{align} which simplifies in the Abelian string case. Therefore the flux of a string is not a unique element in the homotopy group of the manifold, but rather the conjugacy classes of the manifold correspond to the strings. Hence we get the powerful result that elements of the same conjugacy classes must correspond to strings of the same tension as previously pointed out~\cite{Avgoustidis:2007aa, Preskill:1991kd,McGraw:1996py,McGraw:1997nx}. This shows that there are a finite number of tension parameters that each symmetry breaking chain can have, which is related to the number of conjugacy classes. 

\par We cannot generally say that if the first homotopy group of the manifold generated from the spontaneously breaking of a non-Abelian group as having N conjugacy classes, then there are N distinct tensions. This is due to the fact that some of the conjugacy classes contain the inverse elements of another conjugacy class. Therefore those elements correspond to the anti-strings of whatever the inverse conjugacy class is, which will be worked out in detail in Sec. \ref{A4}. Since we have established that strings in the same conjugacy class must have the same tension as well as strings and anti-strings having the same tension, there are in general less tension parameters than conjugacy classes.  Furthermore, the exact form and values of the tensions will depend on the exact field theory that is written down, i.e. number of scalars and representation assignments. For the case of our current work,  we just label each conjugacy class with a benchmark tension value.

\par We will look at the gravitational wave spectrum for the breaking of $\rm SO(3)$ into $\rm A_4$ and investigate a benchmark point of tension to compare the spectrum from $\rm U(1)$ cosmic strings. Our analysis will rely purely on the group theory of $\rm SO(3)$ and $\rm A_4$. We will include the interesting interactions  that can arise in non-Abelian string networks as has been previously described in past literature~\cite{McGraw:1996py,McGraw:1997nx,Alford:1992yx}.

\section{Breaking \texorpdfstring{$SO(3)$}{SO(3)} to Finite Subgroups}
\label{Breaking}
\subsection{Topological Structure of the Manifold}
\label{proof}
Since cosmic strings are classified through conjugacy classes of the first homotopy group of the vacuum manifold, an understanding of the topology of our theory is warranted. This analysis will tell us exactly which group we must study in order to extract string properties. We first note that $\rm SO(3)$ is not simply connected so we cannot make use of the useful relation $\rm \pi_1(G/H)\cong \pi_0(H)$. We will instead look at the double cover of $\rm SO(3)$, which is $\rm SU(2)$, so we have a 2-to-1 homomorphism
\begin{equation}
    \rm p:SU(2)\rightarrow SO(3)\,.
\end{equation} 
By definition we have the preimage $\rm p^{-1}(A_4) \cong T'$, and we have  
\begin{align}
     \rm SU(2)/p^{-1}(A_4)\cong SO(3)/A_4\,.
\end{align} 
Since SU(2) is simply connected, we can use the long exact sequence 
\begin{equation}
    \rm 0=\pi_1(SU(2))\rightarrow \pi_1(SU(2)/T')\rightarrow\pi_0(T')\rightarrow \pi_0(SU(2))=0
\end{equation} 
and therefore $\rm\pi_1(SU(2)/T') \cong \pi_0(T')=T'$. Hence we have \begin{align*}
    \rm \pi_1(SU(2)/T')\cong \pi_1(SO(3)/A_4) 
\end{align*}
and can therefore conclude 
\begin{equation}
    \pi_1\rm(SO(3)/A_4)\cong T' \, ,
\end{equation}
and thus we must study the conjugacy classes of the binary tetrahedral group $\rm T'$. Note that there are more conjugacy classes for $\rm T'$ than there is for $\rm A_4$. However, we will show that the number of ``distinct'' conjugacy classes and thus ``distinct'' strings in $T^{\prime}$ is equal to the number of conjugacy classes of $\rm A_4$, as some classes contain the anti-strings of others. 

\subsection{Breaking \texorpdfstring{$SO(3)$}{SO(3)} to \texorpdfstring{$A_4$}{A4}}
\label{A4}
\par To study the gravitational wave spectrum of the resulting string network from $\rm SO(3)\rightarrow A_4$, we must look into the conjugacy classes of $\rm T^{\prime}$. The conjugacy classes will determine how many distinct string species there are and the group multiplication rules will dictate how the different string species will combine and evolve. The binary tetrahedral group has 7 conjugacy classes, with 6 non-trivial ones corresponding to 6 potential tension parameters. If we identify the string with its own anti-string, then there are only 4 unique tension parameters that go into our analysis as we will now show. 
\par 
A discussion of the properties of $\text{T}'$ is warranted as this will make our discussion clear. We will follow the conventions of $\text{T}'$ in \cite{Ishimori:2010au} with presentation 
\begin{equation}
\label{eq:pres}
\rm{T'} = \langle s, t, r \mid s^2 = r,\ r^2 = t^3 = (st)^3 = e,\ rt = tr \rangle
\end{equation} and conjugacy classes
\begin{align*}
    C_1 &= \{e\} \\ C_1'& = \{r\} \\ C_4 &= \{t,rsts,st,ts \} \\ C_4' &=\{t^2,rtst,rt^2s,rst^2\} \\ C_6&=\{s,rs,tst^2,t^2st,rtst^2,rt^2st\} \\ C_4^{''}&=\{rt,sts,rst,rts\} \\ C_4^{'''}&=\{rt^2,tst,t^2s,st^2\} \; .
\end{align*}
From the presentation, we can derive useful relations among the group elements 
\begin{align*}
    r^{-1}&=r \; , &t^{-1}&=t^2 \; ,\\
    s^{-1}&=rs \; , &(st)^{-1}&=t^{-1}s^{-1}=t^2rs=rt^2s \; ,\\tst&=st^2s \; .
\end{align*}
\par We then search for the inverse elements within each conjugacy class. Since $C_1=\{e\}$ it corresponds to a trivial defect and hence has no cosmic string associated with it. $C_1'=\{r\}$ contains its own inverse, and thus is a unique unoriented string species corresponding to one string tension parameter. For $C_6=\{s,rs,tst^2,t^2st,rtst^2,rt^2st\}$ we have, 
\begin{align*}
    s^{-1}&=rs\in C_6 \; , &(rs)^{-1}&=s \in C_6 \; , \\
    (tst^2)^{-1}&=rtst^2\in C_6 \; , &(rtst^2)^{-1}&=tst^2 \in C_6 \; .
\end{align*}
Therefore $C_6$ is closed under inversion. Thus it corresponds to one string species; two such strings may annihilate if they interact similar to what was discussed in~\cite{McGraw:1996py,McGraw:1997nx}. For $C_4=\{t,rsts,st,ts\}$ we have,
\begin{align*}
    t^{-1}&=t^2\in C_4' \; , &(rsts)^{-1}&=rtst\in C_4' \; , \\
    (st)^{-1}&=rt^2s \in C_4' \; , &(ts)^2&=rt^2s\in C_4' \; .
\end{align*} 
Therefore $C_4$ and $C_4'$ contain each other's inverses, but since they must have the same tension, we use the same tension parameter to describe them and can treat them as the same string in our scenario. For $C_4''=\{rt,sts,rst,rts\}$ we have,
\begin{align*}
    (rt)^{-1}&=rt^2\in C_4''' \; , &(sts)^{-1}&=tst\in C_4''' \; ,\\ 
    (rst)^{-1}&=t^2s\in C_4''' \; , &(rts)^{-1}&=st^2\in C_4''' \; .
\end{align*} 
Thus we see that $C_4''$ and $C_4'''$ contain each other's inverses and have the same tension giving us our fourth and final tension parameter.

\par It is important to distinguish the number of distinct string species and the number of tension parameters. We have previously discussed that the number of string species is equal to the number of conjugacy classes of the first homotopy group whereas the number of tension parameters does not. However, this is not to say that there are 4 string species in this symmetry breaking chain and should not be treated as such, given that in the case of entangled network dynamics strings from different conjugacy classes can interact.
\par We worked out the multiplication of conjugacy classes via Bernside's lemma which states that for conjugacy classes $C_i,\,C_j,\, C_k$ we have $C_iC_j=\sum_{k}a_{ijk}C_k$ where 
\begin{equation}
\label{eq:burnside}
    a_{ijk}=\frac{|C_i||C_j|}{|G|}\sum_{\chi \in \text{irrep}}\frac{\chi(g_i)\chi(g_j)\overline{\chi(g_k)}}{\chi(1)} \; ,
\end{equation}
where $C_i,C_j\in G$, and  $|C_i|,|C_j|,|G|$ are the orders of $C_i,C_j,G$ respectively, and $\chi$ is the character of the irreducible representation of group elements. This allows us to find the multiplicity of each conjugacy class in the product. Applying ~\ref{eq:burnside} we get the following multiplication table via GAP~\cite{GAP4}, 
\begin{table}[h]
\centering
\resizebox{\textwidth}{!}{%
\renewcommand{\arraystretch}{1.3}
\begin{tabular}{c|ccccccc}
$\cdot$ & $C_1$ & $C_1'$ & $C_4$ & $C_4'$ & $C_6$ & $C_4''$ & $C_4'''$ \\
\hline
$C_1$      & $C_1$                    & $C_1'$                   & $C_4$                    & $C_4'$                   & $C_6$                    & $C_4''$                  & $C_4'''$                 \\
$C_1'$     & $C_1'$                   & $C_1$                    & $C_4''$                  & $C_4'''$                 & $C_6$                    & $C_4$                    & $C_4'$                   \\
$C_4$      & $C_4$                    & $C_4''$                  & $C_4' + 3C_4'''$         & $4C_1 + 2C_6$            & $3C_4 + 3C_4''$          & $3C_4' + C_4'''$         & $4C_1' + 2C_6$           \\
$C_4'$     & $C_4'$                   & $C_4'''$                 & $4C_1 + 2C_6$            & $C_4 + 3C_4''$           & $3C_4' + 3C_4'''$        & $4C_1' + 2C_6$           & $3C_4 + C_4''$           \\
$C_6$      & $C_6$                    & $C_6$                    & $3C_4 + 3C_4''$          & $3C_4' + 3C_4'''$        & $6C_1 + 6C_1' + 4C_6$    & $3C_4 + 3C_4''$          & $3C_4' + 3C_4'''$        \\
$C_4''$    & $C_4''$                  & $C_4$                    & $3C_4' + C_4'''$         & $4C_1' + 2C_6$           & $3C_4 + 3C_4''$          & $C_4' + 3C_4'''$         & $4C_1 + 2C_6$            \\
$C_4'''$   & $C_4'''$                 & $C_4'$                   & $4C_1' + 2C_6$           & $3C_4 + C_4''$           & $3C_4' + 3C_4'''$        & $4C_1 + 2C_6$            & $C_4 + 3C_4''$           \\
\end{tabular}%
}
\caption{Class multiplication table for $\rm T'$. Each entry $C_i \cdot C_j = \sum_k a_{ijk} C_k$ gives the class product, with $a_{ijk}$ being the multiplicity of ordered pairs $(g,h) \in C_i \times C_j$ producing each specific element of $C_k$.}
\label{tab:2T_mult}
\end{table}
which as we will see, is connected to the fusion rules of cosmic strings of different species.

\par As stated before, rules of combination are given by the conjugacy class multiplication tables. From this we can derive probabilities that strings of certain conjugacy classes will interact and form a different string of a different conjugacy class. We then can perform an analysis on the loop formation by running a velocity-dependent one-scale (VOS) model~\cite{Avgoustidis:2007aa} and see how this structure affects gravitational wave production.

\section{Non-Abelian VOS Model}
\par In this section, we will briefly go over the non-Abelian VOS model. There has been a lot of literature on this subject in relation to non-Abelian cosmic strings and cosmic superstings~\cite{Avgoustidis:2007aa,Nakamura:2025ptd,Martins:2004vs,Tye:2005fn,Copeland:2005cy,Copeland:2006eh,Saffin:2005cs,Hindmarsh:2006qn,Cella:2015btn}. The VOS model is used to determine the properties of long cosmic string segments. It tells us how the energy density $\rho$ and root mean square (rms) velocity $v$ of the cosmic strings evolve with time. From this, we can deduce the loop populations and formations of different string types, as well as encode the information of interactions between strings. We must run our network through this model to then determine the gravitational wave radiated by loops. A key feature of the VOS model is its prediction of a scaling solution where the characteristic length scale is growing proportionally to cosmic time leading to a stable solution to the equations. 
\par In the Abelian case, we start with defining the energy density of the string network as
\begin{equation}
     \rho =\frac{\mu}{L^2} \; ,
\end{equation} 
where $\mu$ is the string tension and L is the characteristic length of the string network. From the energy momentum tensor one is able to get the following equation governing the energy density of the string network \begin{equation}
   \dot{\rho}=-2\frac{\dot{a}}{a}(1+v^2)\rho-\frac{\tilde{c}v\rho}{L} \; ,
\end{equation}
where $a$ is the scale factor and $\tilde{c}$ is the loop production parameter. 
From the Nambu-Goto action one can get the equation for the rms velocity of string segments
\begin{equation}
    \dot{v}=(1-v^2)\left(\frac{k}{R}-2\frac{\dot{a}}{a}v\right)\; ,
\end{equation} 
 with $R$  being the typical radius of curvature of the strings in the network. The momentum parameter~\cite{Martins:1995tg} $k$ is taken to be 
\begin{equation}
    k=k(v)=\frac{2\sqrt{2}}{\pi}\left( \frac{1-8v^6}{1+8v^6}\right)
\end{equation} 
given in~\cite{Martins:2000cs}. Taking $R=L\equiv\gamma(t)t$, where $\gamma$ is the dimensionless correlation length, we then have 
\begin{equation}
    \frac{\dot{\gamma}}{\gamma}=\frac{1}{2t}\left(2Ht(1+v^2)-2+\frac{\tilde{c}v}{\gamma}\right) \; ,
\end{equation}
where $H$ is the Hubble parameter.

\par In the non-Abelian case, these equations must be modified since strings cannot only pass through each other or reconnect, which only occurs if strings of the same type interact. There are topologically excluded outcomes due to the conservation of flux. We then have four topologically allowed outcomes. Consider two strings of flux $a$ and $b$ respectively. Remembering that $a$ and $b$ are group elements of $\pi_1(\mathcal{M})$ if the group commutator \begin{equation}
    [a,b]\equiv aba^{-1}b^{-1}
    \label{commutator}
\end{equation} is the identity element $e$, i.e. the elements commute, then the two strings will pass through one another. If $a=b$ or $a=b^{-1}$ then we get reconnection, else the two strings will form a junction with flux $aba^{-1}b^{-1}$~\cite{Vilenkin:2000jqa} which is called a bridge. We also have one other topologically allowed outcome referred to as zippers~\cite{Avgoustidis:2007aa} in which two strings coalesce along their lengths. Since zippering relies on two strings having close range interaction, meaning a low energy shallow collision, we will not be considering zippers in our analysis. However they can have significant effects on the time in which the network achieves a scaling solution~\cite{Nakamura:2025ptd}.
The derivation has been done in great detail previously~\cite{Avgoustidis:2007aa} so we will quote the results and point out our modification. The resulting equations, excluding zippers, are \begin{align}
    \frac{\dot{\gamma}_i}{\gamma_i}&=\frac{1}{2t}\left[2Ht(1+v_i^2)-2+\frac{\tilde{c}_iv_i}{\gamma_i}-\sum_{j,k\leq j}\tilde{d}^i_{jk}\bar{v}_{jk}\frac{\gamma_i^2}{\gamma_j^2\gamma_k^2}\frac{\ell^i_{jk}(t)}{t}\right] \; , \\
    \dot{v}_i&=\frac{1-v_i^2}{t}\left[ \frac{k_i}{\gamma_i}-2Ht v_i-\sum_{j,k} \tilde{d}^k_{ij}w_i(\mu_i,\mu_j)\frac{\bar{v}_{ij}}{v_i}\frac{\mu_k}{\mu_i}\frac{1}{\gamma_j^2}\frac{\ell_{ij}^k}{t}\right] \; .
\end{align}
Here 
\begin{align*}
    \gamma_i&= \frac{L_i}{t} \; , \\
    \bar{v}_{ij}&=\sqrt{v_i^2+v_j^2} \; ,\\
    \frac{\ell_{ij}^k}{t}&=\frac{\mu_i\gamma_iv_i^2+\mu_j\gamma_jv_j^2}{2\mu_k} \; ,\\
    w(\mu_i,\mu_j)&=\frac{\mu_j}{\mu_i+\mu_j} \; ,
\end{align*} 
and the coefficients $\tilde{d}_{ij}^k$ is the probability that string $i$ and string $j$ will form a bridge of string $k$ multiplied by some efficiency factor taking energetics into account. Importantly, this quantity $\tilde{d}_{ij}^k$ has been taken to be a free parameter that can be varied such that a scaling solution is met and there is agreement with numerical simulations. In our case, we posit that one can get at least part of it from the group theory and topological analysis of the symmetry breaking. As such, it is useful to write 
\begin{equation}
    \tilde{d}_{ij}^k=P^k_{ij}\tilde{d} \; ,
\end{equation} where $P_{ij}^k$ is the pure probability we can extract from the group theory and class multiplication in Sec. \ref{Breaking}, and $\tilde{d}$ encodes the kinematic properties such as the angle of collision, which will affect the overall probability. We also have 
\begin{equation}
\label{eq:def_c}
    \tilde{c}_i=P_{i}\tilde{c}
\end{equation} 
where $P_{i}$ is the reconnection probability and $\tilde{c}$ is the efficiency parameter encoding kinematic properties. Once we have the long string network's behavior described by the non-Abelian VOS, we can calculate the gravitational wave spectrum produced by the cosmic strings. First, however, we will demonstrate how to explicitly calculate $P^k_{ij}$ from group theory in the framework of $\rm SO(3) \rightarrow A_4$.
 
\subsection{Probabilities From Class Multiplication} \label{probabilites}
It is important to recall the main points we previously discussed in \ref{A4}. First, we showed that there are four distinct tensions corresponding to 4 distinct string networks where we defined the networks in table \ref{tab:2T_centralizers}. In order to discuss whether strings will pass through one another, reconnect, or form bridges, we will need to look at the centralizer of a representative element, i.e. all elements in $\rm T'$ that commutes with said element. 
\begin{table}[h]
\centering
\renewcommand{\arraystretch}{1.3}
\begin{tabular}{c|c|c}
Network & Representative & Centralizer \\
\hline
$1 \;\; (C_6)$                 & $s$  & $\{e,\, s,\, r,\, rs\}$ \\
$2 \;\; (C_1')$                & $r$  & $\rm T'$ \\
$3 \;\; (C_4'' \cup C_4''')$   & $rt$ & $\{e,\, rt,\, t^2,\, r,\, t,\, rt^2\}$ \\
$4 \;\; (C_4 \cup C_4')$       & $t$  & $\{e,\, t,\, t^2,\, r,\, rt,\, rt^2\}$ \\
\end{tabular}
\caption{String networks of $\rm T'$, a representative flux for each, and its centralizer.}
\label{tab:2T_centralizers}
\end{table}
\par We will denote an interaction of a string from network $i$ and a string of network $j$ interacting as $(i,j)$. For the $(1,1)$ case taking representative $s$ we have $C_{\rm T'}(s)=\{e, s, r, rs\}$ and $C_6=\{s,rs,tst^2,rtst^2,rt^2st\}$. Since $s,\, rs \in C_{\rm T'}(s)$ they will reconnect when two strings of those fluxes interact. Therefore given two strings from string network 1 interact, assuming strings of all fluxes are created equally, there is a $P_{\rm recon}=\frac{2}{6}=\frac{1}{3}$ probability of reconnection. For bridges, there are two options that can happen as can be seen from table \ref{tab:2T_mult}, $C_6\times C_6=6C_1+6C_1'+4C_6$ either the resulting flux ends up in $C_1'$ or $C_6$. To figure out which one, we must look at the group commutator given in Eq. \ref{commutator}. We will compute the commutator for each possible interaction of flux. 
\begin{align*}
    [s,tst^2]&=r \in C_1' \; ,\quad\quad [s,rtst^2]=r\in C_1' \; ,\\ 
    [s,t^2st]&=r \in C_1' \; , \quad \quad [s,rt^2st]=r \in C_1' \; ,
\end{align*}
hence all other interactions land in $C_1'$ therefore $P_{11}^2=\frac{2}{3}$. Notice we are able to just do one representative because each element within a string network is related by conjugation and we have $a'=gag^{-1}$ and $b'=gbg^{-1}$
\begin{align*}
    [a',b']&=a'b'(a^{-1})'(b^{-1})' \\&=gaba^{-1}b^{-1}g^{-1} 
    \\&=g[a,b]g^{-1}
\end{align*}
therefore the fluxes are in the same conjugacy class. Interestingly, since network 2 has a single string with flux $r$ which commutes with the entirety of $\rm T'$, it will pass through every string except for another network 2 string in which it will reconnect with probability 1 so $P_{\rm recon}^2=1$. The explicit calculations are tedious so we refer the reader to Appendix~\ref{appendixA}.We list the probabilities of certain string interactions in Table \ref{tab:2T_bridges}. 
\begin{table}[h]
\centering
\renewcommand{\arraystretch}{1.3}
\begin{tabular}{c|cccc}
Pair & Reconnect & Pass through & Bridge & Bridge lands in \\
\hline
$(1,1)$ & $1/3$ & $0$   & $2/3$ & Network 2 \\
$(1,2)$ & $0$   & $1$   & $0$   & Passes Through    \\
$(1,3)$ & $0$   & $0$   & $1$   & Network 1 \\
$(1,4)$ & $0$   & $0$   & $1$   & Network 1 \\
$(2,2)$ & $1$   & $0$   & $0$   & Reconnects      \\
$(2,3)$ & $0$   & $1$   & $0$   & Passes Through      \\
$(2,4)$ & $0$   & $1$   & $0$   & Passes Through      \\
$(3,3)$ & $1/4$ & $0$   & $3/4$ & Network 1 \\
$(3,4)$ & $0$   & $1/4$ & $3/4$ & Network 1 \\
$(4,4)$ & $1/4$ & $0$   & $3/4$ & Network 1 \\
\end{tabular}
\caption{Collision outcomes for the string networks of $\rm T'$ with bridge-type junctions. Each entry is the fraction of ordered flux pairs $(a,b)$, with $a$ in the first network and $b$ in the second, that reconnect ($b = a^{\pm 1}$), pass through ($ab = ba$, $b \neq a^{\pm 1}$), or form a bridge carrying $aba^{-1}b^{-1}$ ($ab \neq ba$). The last column gives the network containing the bridge flux.}
\label{tab:2T_bridges}
\end{table}

\subsection{Dimensionless Correlation Length and Energy Density Evolution}
In this section, we will present the numerical results of the non-Abelian VOS equations. We demonstrate that each string network attains a scaling regime, with the dimensionless correlation length $\gamma = L/t$, the energy density $\rho$, and the rms velocity $v_{\rm rms}$ all approaching scaling behavior. We show these results for limiting values of bridging starting from no bridging $\tilde{d}=0$, to maximal bridging, $\tilde{d}=\tilde{c}$. Additionally, we compare the evolution of these quantities to a standard $\rm U(1)$ cosmic string network, thus showing how bridging affects the evolution of the network.

\par
\begin{figure}[h]
    \centering
    \includegraphics[width=\textwidth]{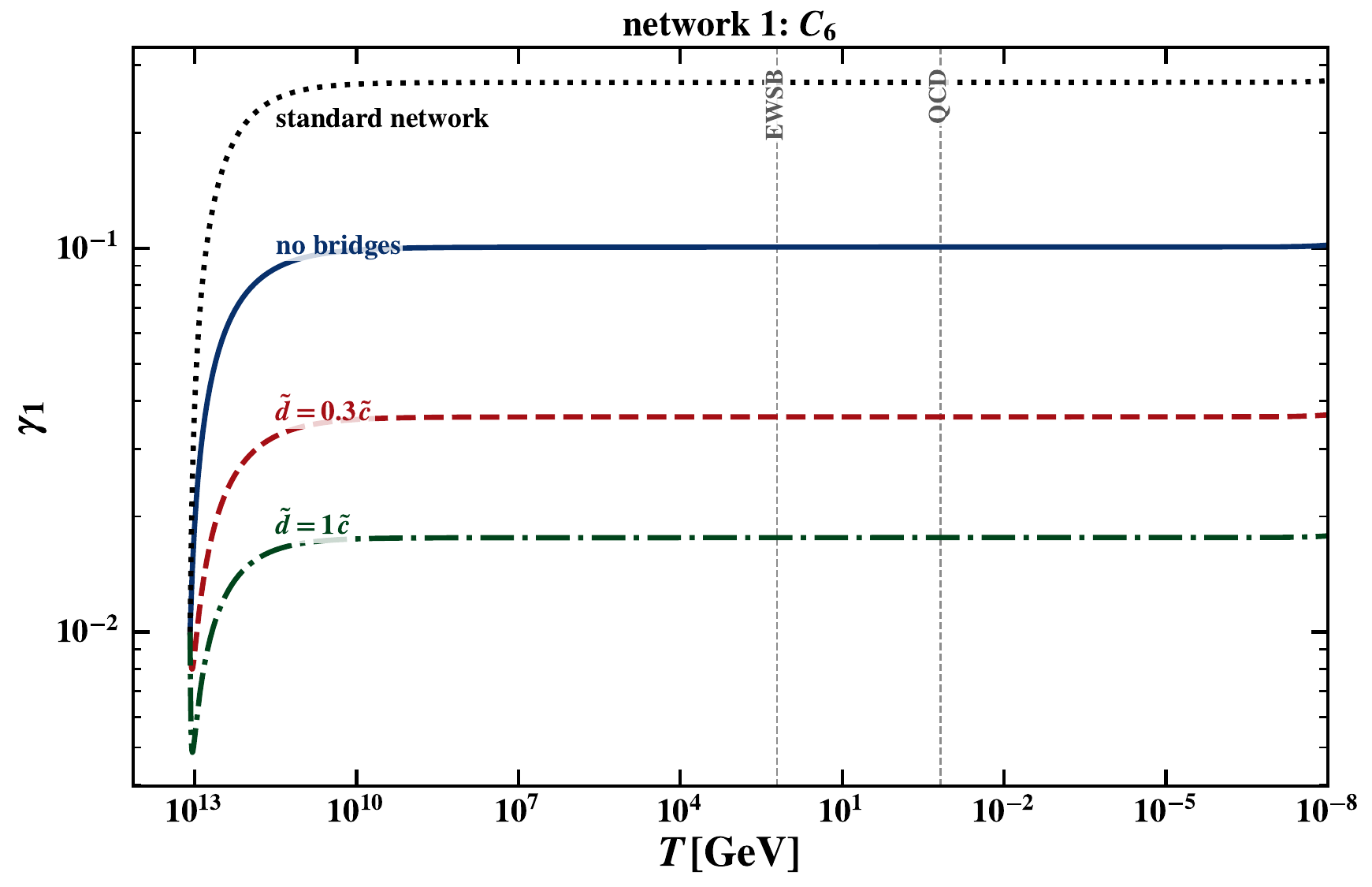}
    \caption{ Evolution of dimensionless correlation length $\gamma_{1}$ with no bridging and bridging taken into affect for Network 1 compared to the standard Abelian $\rm U(1)$ network.}
    \label{fig:gamma1}
\end{figure}

\begin{figure}[h]
    \centering
    \includegraphics[width=\textwidth]{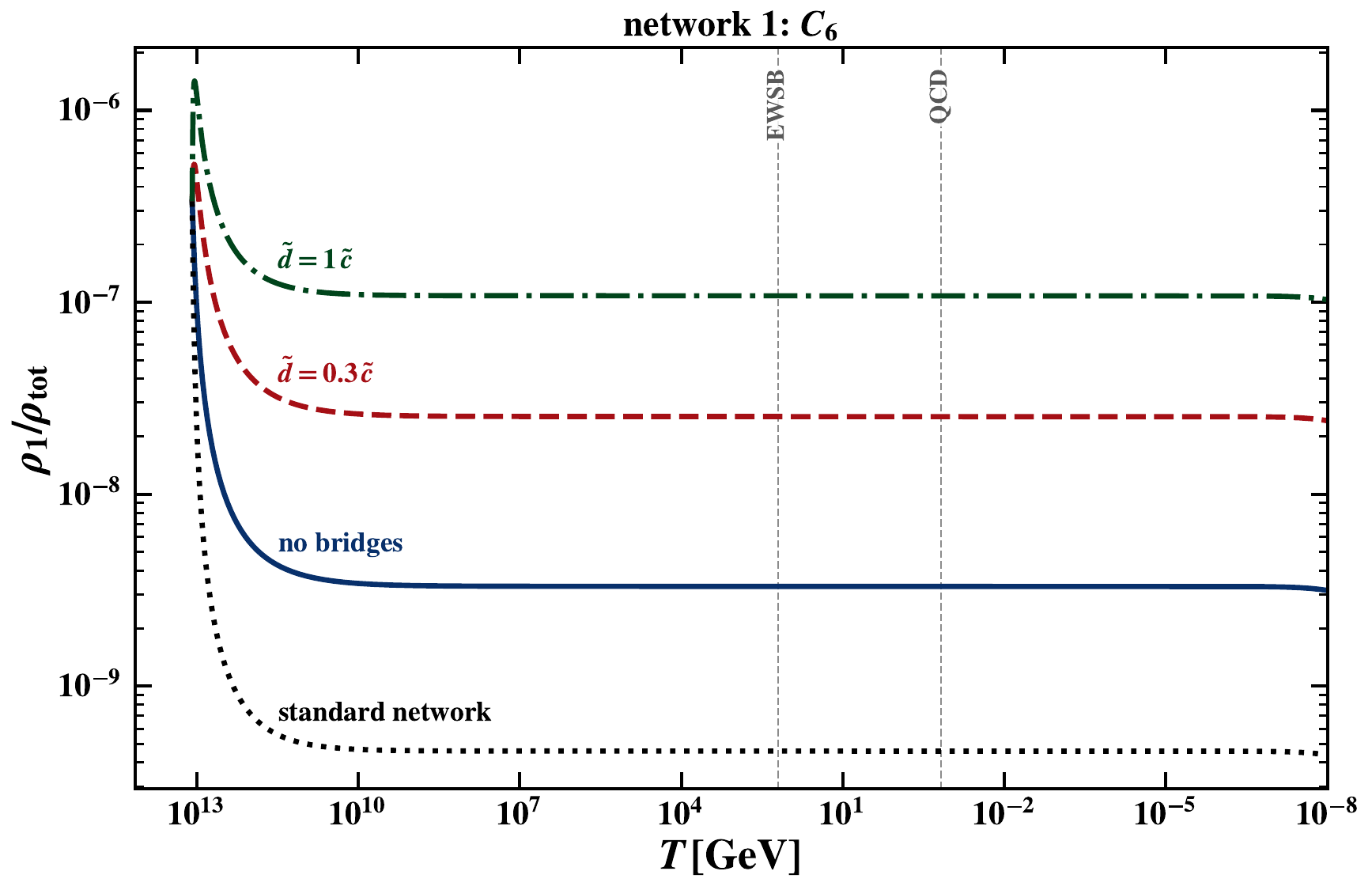}
    \caption{ Evolution of the energy density with no bridging and bridging taken into affect for Network 1 compared to the standard Abelian $\rm U(1)$ network.}
    \label{fig:rho_1}
\end{figure}

 For the time being, we will illustrate the effects for string network 1 and 2; the other cases will be shown in appendix \ref{appendixB}. For the case of network 1, where a string interacts with itself, we have a reconnection probability of $\frac{1}{3}$ and a bridging probability of $\frac{2}{3}$ with the resulting bridge landing in network 2. Since all network interactions besides the (1,2) interaction leads to bridging in network 1, we have an increase in the production of network 1 strings. Further, as the rate of bridging is larger than the rate of reconnection for networks 3 and 4, there is an increased energy density of network 1 and we therefore expect a smaller correlation length compared to the standard Abelian network. This can be seen in Fig. \ref{fig:gamma1} .
We see as we increase the bridging efficiency parameter $\tilde{c}$ that contains kinematic information, the correlation length drops even more. Note that there is still a decrease in correlation length with no bridging because the reconnection probability is altered by the group structure of $\pi_1(\mathcal{M})$. Additionally one can understand the increase in correlation length at the right side of the plot as the moment when loop production increases and subsequently radiates GW thereby diluting the cosmic string network. Further we show the resulting energy density evolution of network 1 in Fig. \ref{fig:rho_1}. There is an increase in energy density compared to the standard $\rm U(1)$ network, as expected from the evolution of the correlation length. 

\par
\begin{figure}[h]
    \centering
    \includegraphics[width=\textwidth]{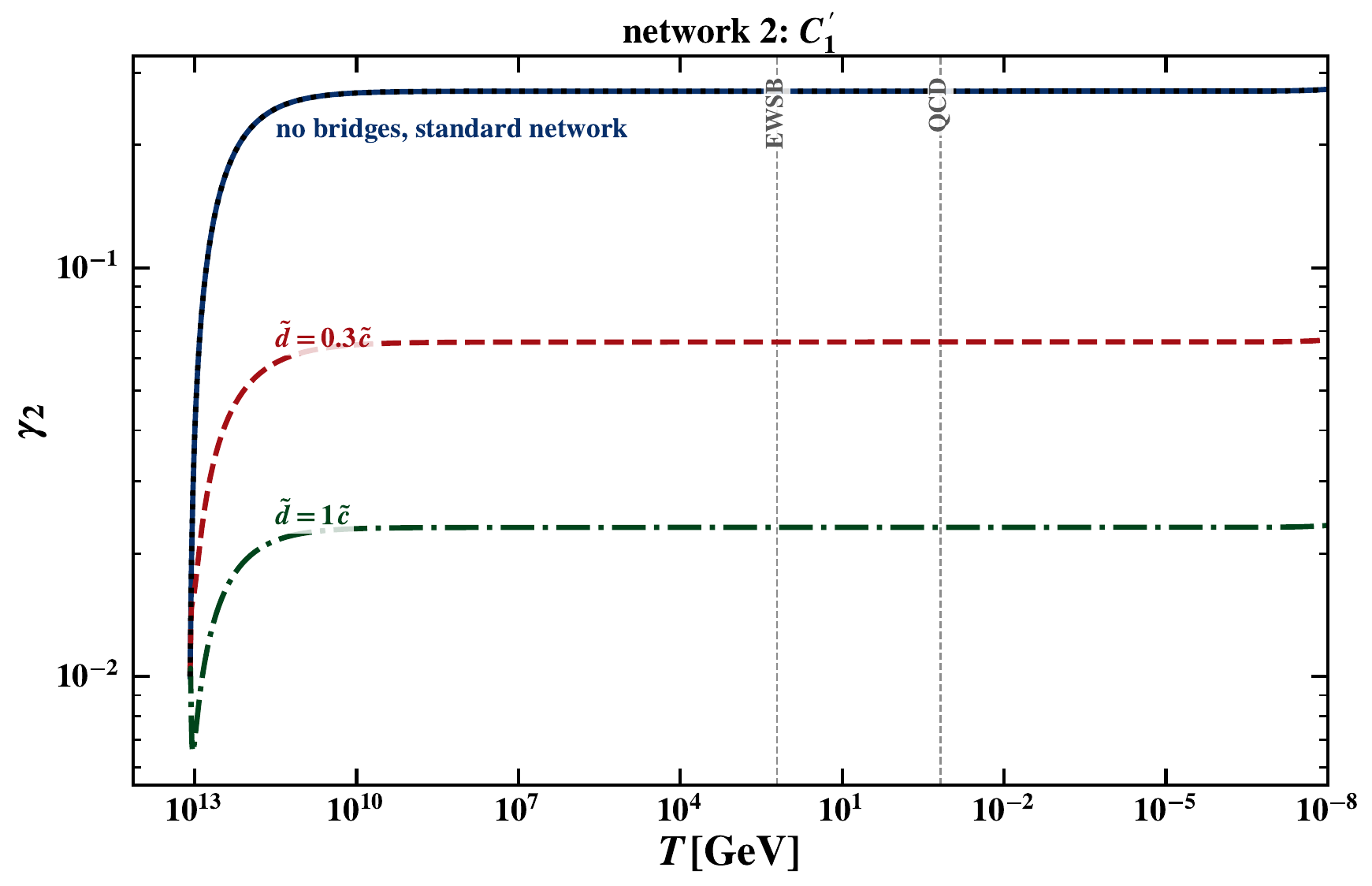}
    \caption{Correlation length evolution for network 2.}
    \label{fig:gamma_2}
\end{figure}

\begin{figure}[h]
    \centering
    \includegraphics[width=\textwidth]{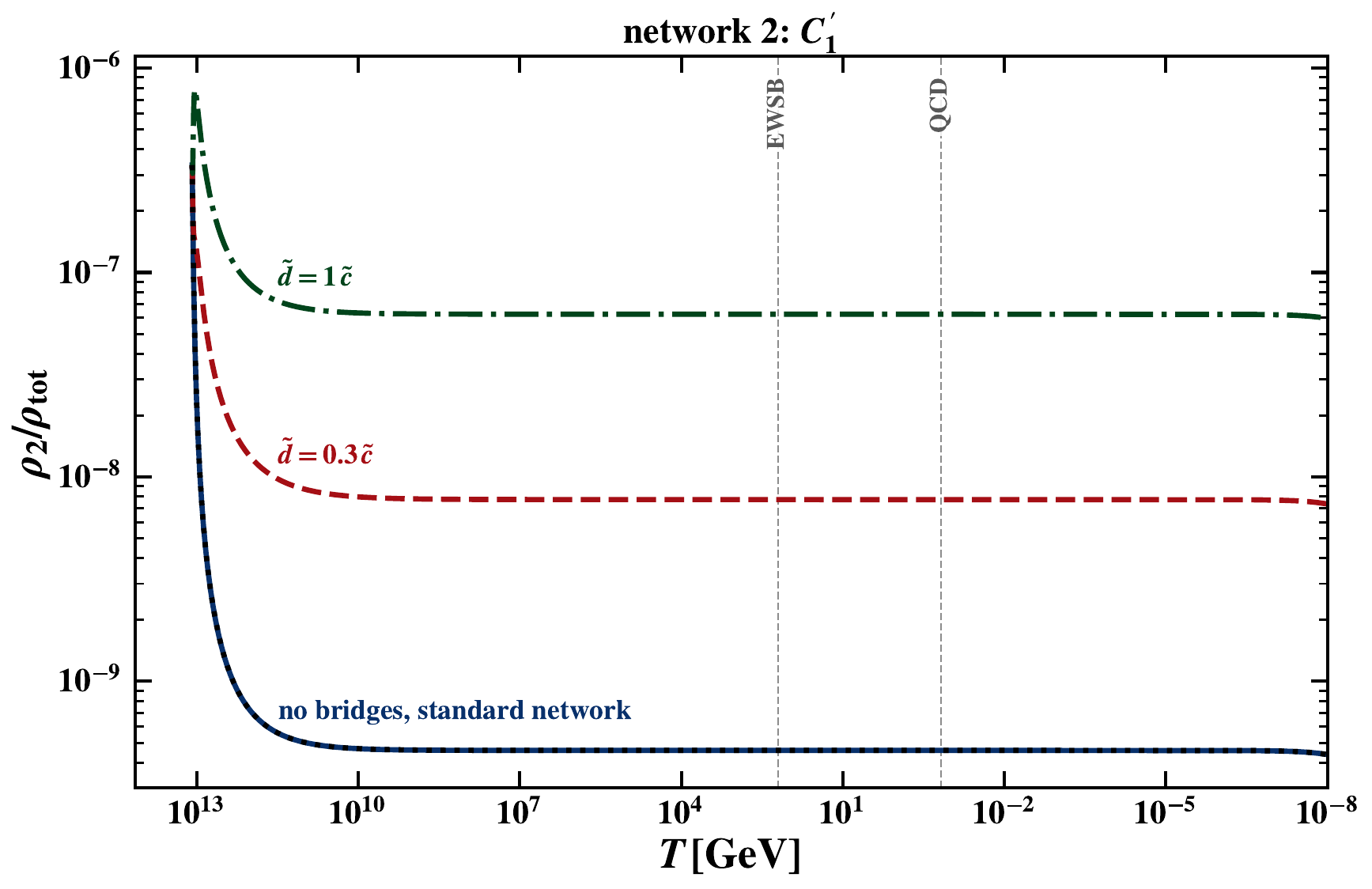}
    \caption{Energy density for network 2.}
    \label{fig:rho_2}
\end{figure}

 It is worthwhile to show how the correlation length and the energy density differs across networks since there are different coefficients in the VOS equations dictated by the group structure. Since network 2 will only reconnect and hence has enhanced loop formation, it serves as a good metric in which to compare with other networks. As reconnection is the only possibility for network 2 interactions with itself, it behaves very similar to the Abelian case, and in the limit of no bridging, behaves exactly like the Abelian case as seen in Fig. \ref{fig:gamma_2}, with complimentary energy density seen in Fig. \ref{fig:rho_2}. As different values of the bridging efficiency parameter are scanned, the correlation length for network 2 is larger than that of network 1 which is due to the Abelian nature of network 2. From a group theory point of view, this can be understood since $C_1'=\{r\}$, which is the centralizer of $\rm T'$; thus it commutes with all group elements of $\rm T'$ and is its own inverse. This demonstrates that the topological structure of the vacuum manifold is very important in the dynamics and evolution of cosmic strings. Having established the evolution of the long-string network, we now turn to the resulting gravitational wave spectrum.

\section{Gravitational Waves From Non-Abelian Networks}
\subsection{Loop Density and Gravitational Wave Emission}
\par Before showing the results from our calculation, it is important to go over the theory of gravitational waves from cosmic strings. We adopt an approach given by~\cite{Gouttenoire:2019kij}. The emission of gravitational waves from cosmic strings predominantly comes from the decay of string loops. In order to calculate the gravitational wave  spectrum, we thus need to know how loops form and the rate at which the long-string network loses energy to loop production. For the long-string network $i$, we have 
\begin{equation}
    \dot{\rho}_i=-2H(1+v^2)\rho_i-\dot{\rho}_i|_{\rm loop} 
    \; .
\end{equation}
We define~\cite{Vilenkin:2000jqa} \begin{align*}
    \dot{\rho}_i|_{\rm loop}\equiv \sqrt{2}\frac{\mu}{t^3}\tilde{C}_{\rm eff}=\sqrt{2}\frac{\mu}{t^3}\frac{P_{ii}\tilde{c}_iv_i}{\gamma_i^3} \; ,
\end{align*}
where $P_{ii}$ can be found in Table \ref{tab:2T_bridges}, and $\tilde{c}$ is the efficiency parameter taking kinematics into account defined in Eq. \ref{eq:def_c}. Motivated by the numerical studies~\cite{Blanco-Pillado:2013qja}, we assume loops are formed with a characteristic length $\alpha t$, with $\alpha = 0.1$, and that a fraction $\mathcal{F}_{\alpha}=0.1$ of the energy released by the long-string network goes into such loops. The number density of loops produced per unit time is then
\begin{equation}
    \frac{dn_i}{dt}=\frac{\mathcal{F}_{\alpha}}{\alpha}\frac{P_{ii}\tilde{c}_iv_i}{\gamma_i^3t^4} \; .
    \label{eq:numdens}
\end{equation}
From this point, we can look at the form of the gravitational wave spectrum. Since the gravitational waves are emitted early in the universe, to get the current spectrum we must redshift the spectrum that was previously emitted. To do this, note that gravitational waves redshift as radiation and we therefore have the energy density of emitted gravitational waves as \begin{equation}
    \frac{d\rho_{\rm GW}}{df}=\frac{d\rho_{\rm GW}(\tilde{f},\tilde{t})}{d\tilde{f}}\left( \frac{a(\tilde{t})}{a(t_0)}\right)^3 \; , 
\end{equation} 
where $\tilde{f}=\biggl(\dfrac{a(t_0)}{a(\tilde{t})}\biggr)f$.
After the time of loop formation $t_{\ell}$, we can calculate the decrease of loop length $l_{i}(t)$ as gravitational waves are emitted. We have\begin{equation}
    \frac{dl_i}{dt}=-\Gamma G\mu_i \implies l_{i}(t)=\alpha t_{\ell}-\Gamma G \mu_i(t-t_{\ell}) \; ,
\end{equation} 
where $\Gamma \approx50$~\cite{Blanco-Pillado:2017rnf} and $G\mu_i$ is the string tension. Gravitational waves  emission frequencies are dictated by the normal modes of the string loops,
\begin{align*}
    \tilde{f}=\frac{2k}{l} \; ,
\end{align*} 
where $k$ is a positive integer. The power emitted by a loop of string with mode $k$ is given by 
\begin{equation}
    P_i^{(k)}=\frac{dE_i^{(k)}}{dt}=\Gamma^{(k)}G\mu^2_i \;, \quad  \mathrm{where}\ \sum_k \Gamma^{(k)}=\Gamma
\end{equation} 
and $\Gamma^{(k)}=\frac{\Gamma k^{-4/3}}{3.6}$; therefore the energy density is given by 
\begin{equation}
   \frac{d\rho_{\mathrm{GW}}(\tilde{f},\tilde{t})}{d\tilde{f}} =\int_{t_{\ell}}^{t_0}\frac{dE_{\mathrm{GW}}}{d\tilde{t}}\frac{dn(\tilde{f},\tilde{t})}{d\tilde{f}} \; .
\end{equation} 
Recalling Eq. \ref{eq:numdens} we have \begin{align}
    \frac{dn(\tilde{f},\tilde{t})}{d\tilde{f}}&=\left[\frac{a(t_{\ell})}{a(\tilde{t})}\right]^3\frac{dn}{dt}\frac{dt}{dl}\frac{dl}{d\tilde{f}}\nonumber \\ &=\left[\frac{a(t_{\ell})}{a(\tilde{t})}\right]^3 \sum_k(0.1)\frac{P_{ii}\tilde{c}_iv_i}{\gamma_i^3t_{\ell}}\frac{1}{\alpha(\alpha + \Gamma G \mu_i)}\frac{2k}{f^2}\left[ \frac{a(\tilde{t})}{a(t_0)}\right]^2 \; .
\end{align} 
Using 
\begin{equation}
    \Omega_{\mathrm{GW}}(f)=\frac{8\pi G}{H^2f}\left|\frac{d\rho_{\rm GW}(f,t_0)}{df}\right|
\end{equation} 
we obtain the gravitational wave spectrum from all modes \begin{align}
   &\Omega_{\mathrm{GW}}(f)_i=\sum_k \Omega_{\rm GW}^{(k)}(f)_i \nonumber \\
   &=\sum_k\frac{8\pi G}{H^2}\frac{2k}{f}\frac{\mathcal{F}_{\alpha}\Gamma^{(k)}G\mu_i^2}{\alpha(\alpha+\Gamma G \mu_i)}\int_{t_p}^{t_0}d\tilde{t}\frac{P_{ii}\tilde{c}_iv_i}{\gamma_i^3t_{\ell}^4}\left[ \frac{a(\tilde{t})}{a(t_0)}\right]^5 \left[ \frac{a(t_{\ell})}{a(\tilde{t})}\right] \; ,
\end{align}
where $t_p$ is the time of string formation and $t_{\ell}$ is the time of loop formation given by \begin{equation}
    t_{\ell}(f,\tilde{t})=\frac{1}{\alpha+\Gamma G\mu_i}\left[ \frac{2k}{f}\frac{a(\tilde{t})}{a(t_0)}+\Gamma G \mu_i \tilde{t}\right] \; .
\end{equation}

\subsection{Gravitational Wave Spectra}

\par We calculate the gravitational wave spectrum numerically for a universal string tension of $G\mu=10^{-12}$ across all networks. In principle the tensions of the networks can be different, it depends on the explicit model that is set up, i.e. Higgs representation and potential. Nonetheless it would be interesting to apply this idea to a specific model to see what constraints we can get on parameters in the potential or string tension, we leave such an analysis for future work. We present the results of networks 1 and 2 for the reason that all networks (except for 1 and 2) bridge into network 1, and network 2 only reconnects which is similar to the Abelian string. Further, we present the combined gravitational wave spectrum from all networks and compare all cases to a standard $\rm U(1)$ network.

\par
\begin{figure}[htb!]
    \centering
    \includegraphics[width=\textwidth]{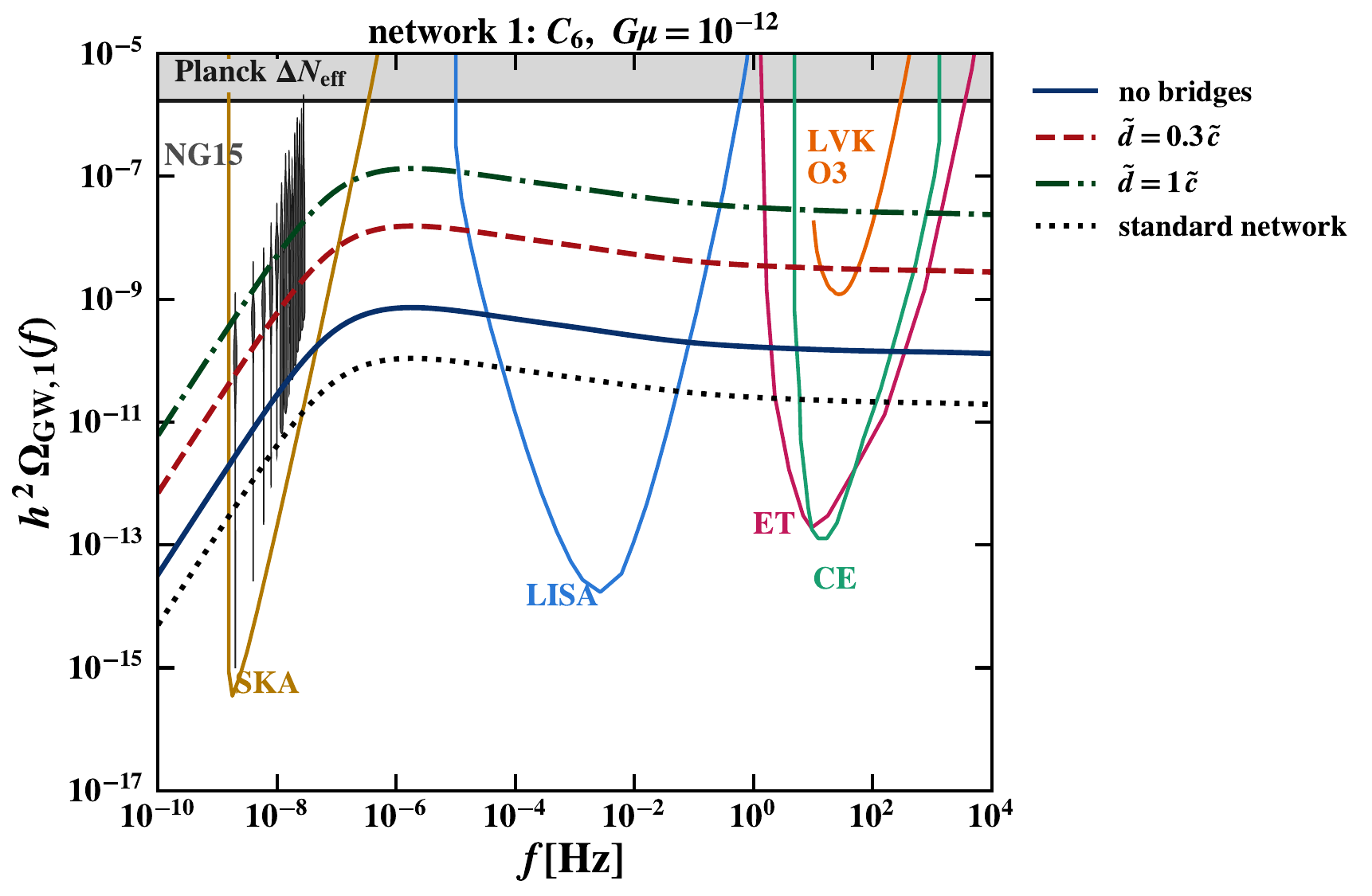}
     \caption{GW spectrum from network 1.}
    \label{fig:gw_1}
\end{figure}

 The result for network 1 is shown in Fig. \ref{fig:gw_1}.  We see that the amplitude of the gravitational wave spectrum depends heavily on the bridging efficiency and probability. This is highly evident for network 1 since all bridges, besides from network 1 and 2, land in network 1. Additionally there is an enhancement in the case of no bridges since the reconnection probability leads to a denser network of loops as can be attested by the total energy density in the network shown in Figs. \ref{fig:gamma1} and \ref{fig:rho_1}.

\par
\begin{figure}[htb!]
    \centering
    \includegraphics[width=\textwidth]{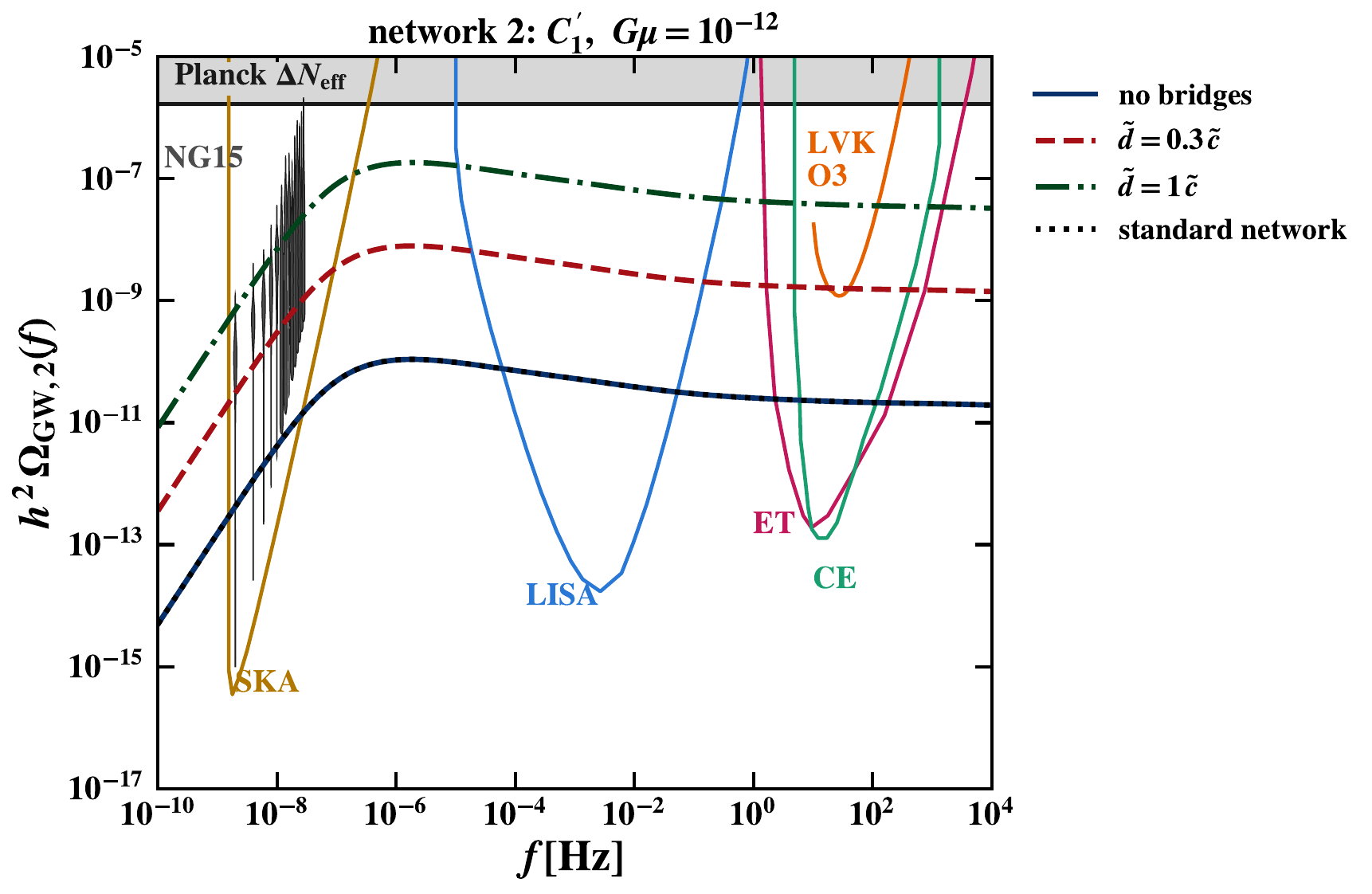}
    \caption{GW spectrum from network 2. In the case of no bridges, the spectrum is identical to that of the standard Abelian network}.
    \label{fig:gw2}
\end{figure}

 We will now compare this to the case of network 2 which behaves very similar to the Abelian case as seen in Fig. \ref{fig:gw2}.
We see that in the limit of no bridges, network 2 behaves exactly like an Abelian string network. This is because a string of network 2 can only reconnect with itself and without production of network 2 strings from bridging, and thus there is no enhancement. This is also understood from the fact that the element corresponding to a string from network 2 commutes with the entirety of $\rm T'$. 

\par
\begin{figure}[htb!]
    \centering
    \includegraphics[width=\textwidth]{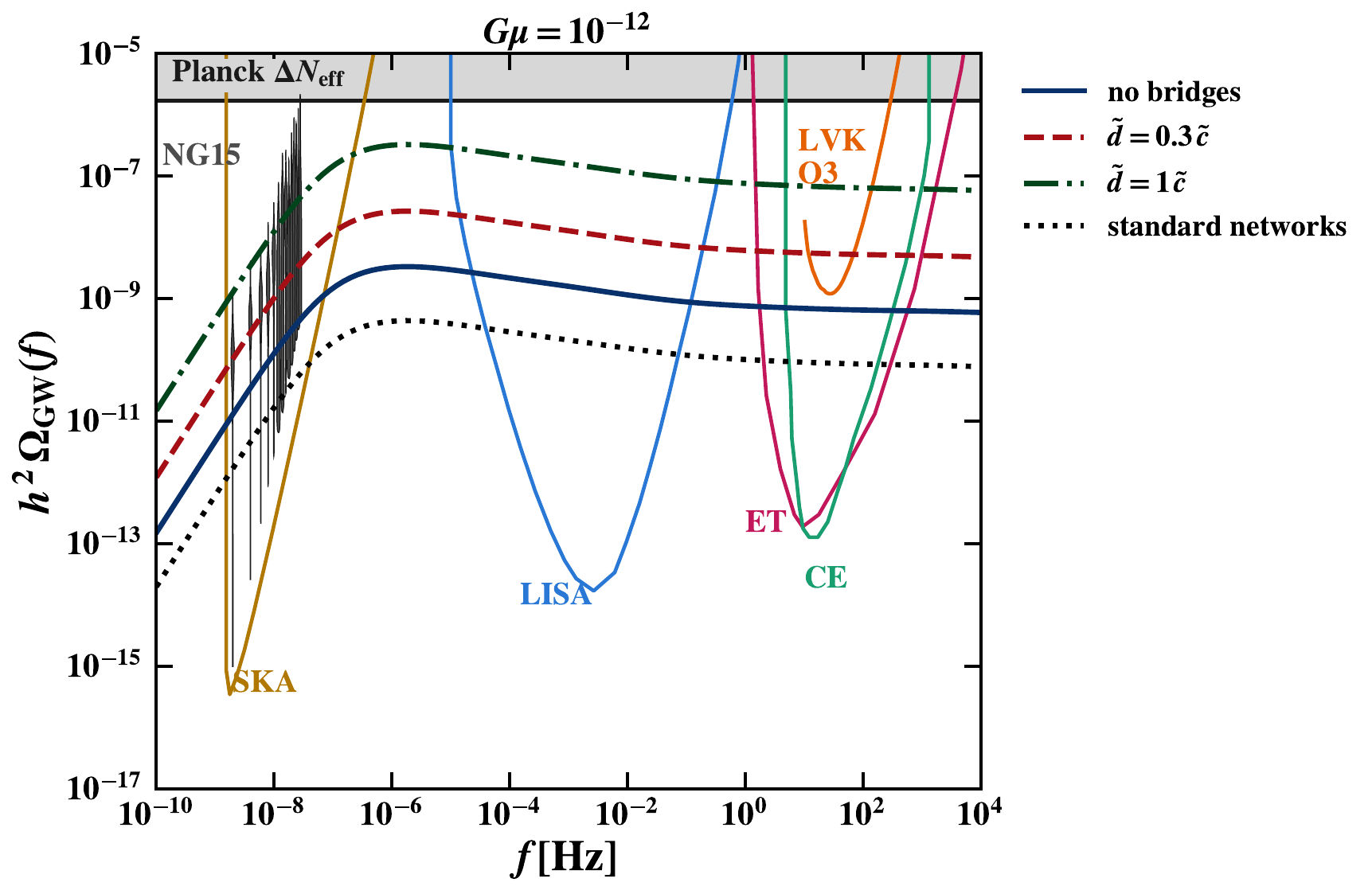}
    \caption{Total gravitational wave signal compared to a U(1) signal. We also see the affect of bridges. The bridges cause an increase in loop number density and hence why we have an increase in GW. There is also a modified probability of reconnection calculated from group theory.} 
    \label{fig:gw_tot}
\end{figure}

 Fig.~\ref{fig:gw_tot} shows the total gravitational wave spectrum due to all four networks. Given the modifications of the bridging and reconnection probabilities, there is an increase in the total gravitational wave spectrum compared to the sum of four individual $\rm U(1)$ networks. It is interesting that due to the non-Abelian nature of the strings, the gravitational wave spectrum will differ so much. This  demonstrates that a theory that breaks from $\rm SO(3) \rightarrow A_4$ is more gravitationally sensitive, compared to a standard $\rm U(1)$ breaking,  assuming the tensions are the same. 

\par
\begin{figure}[h]
    \centering
    \includegraphics[width=\textwidth]{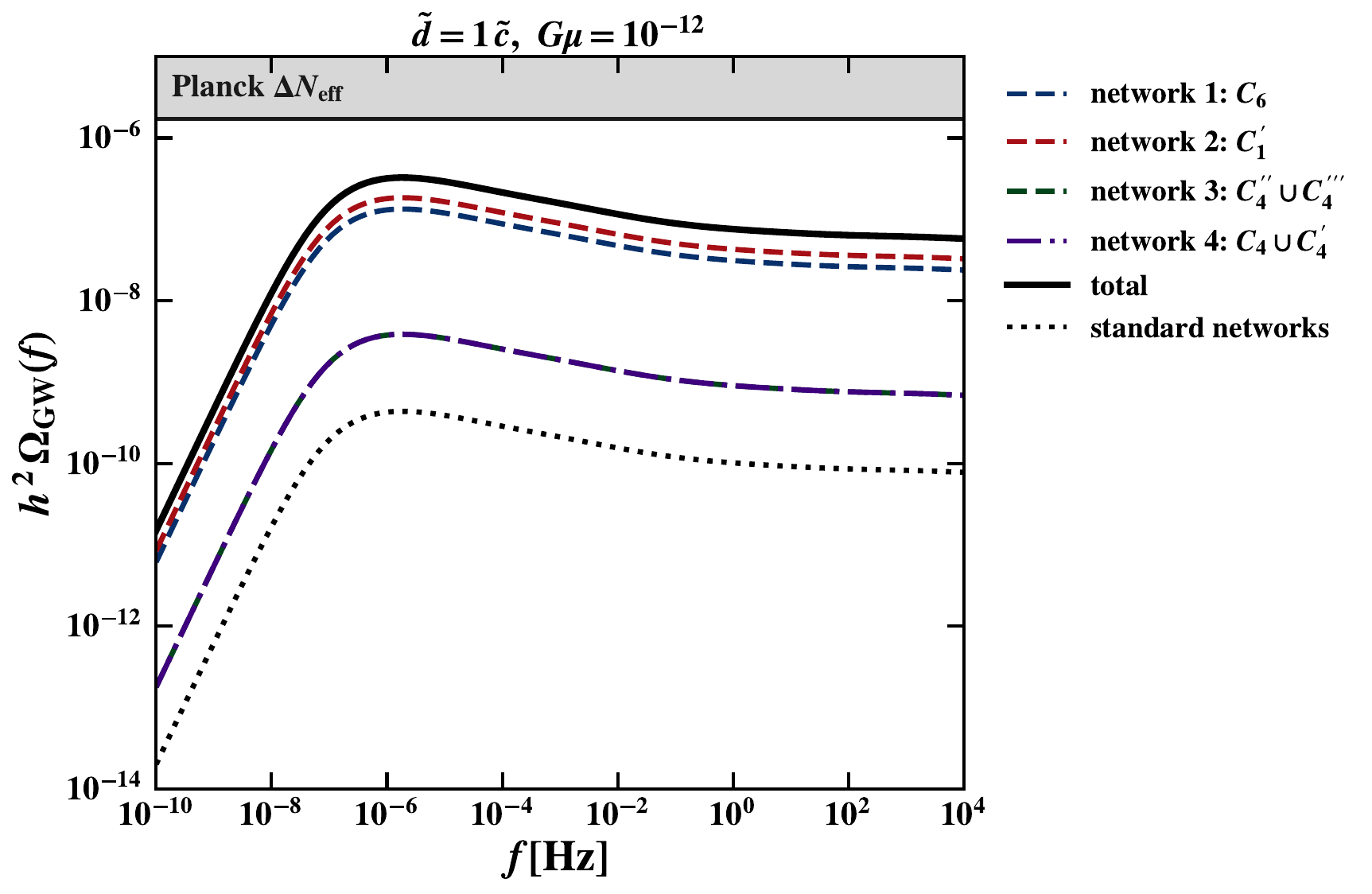}
    \caption{Breakdown of contributions from individual networks to the GW spectrum.}
    \label{fig:breakdown}
\end{figure}

 It is also interesting to note the individual contributions from each network in the case where $\tilde{d}=\tilde{c}$, as shown in Fig. \ref{fig:breakdown}.
It is interesting to note that the dominant contribution of the gravitational wave spectrum are from networks 1 and 2. This is due to the group structure of our theory. Since all string interactions that branch end up in either network 1 or network 2, we have an enhancement of strings of those populations. Therefore it makes sense that the dominant gravitational wave production comes from networks 1 and 2. Note that if we chose a different symmetry to break to, the results will be different. Hence we see a clear relationship between the algebra that characterizes $\pi_1(\mathcal{M})$, and the resulting gravitational wave spectrum. This thus shows that the gravitational wave background produced by cosmic strings is sensitive to the symmetry breaking which occurred. Gravitational wave spectrum thus provides a way to distinguish  flavor models.  
\section{Conclusion}
\par 
In this paper, we analyzed the behavior of cosmic strings that result from the spontaneous symmetry breaking of $\rm SO(3) \rightarrow A_4$, comparing it to the usual $\rm U(1)$ Abelian string network. We then discussed how in the non-Abelian case, the flux of a species is an ill-defined identifier since the defects cannot in general pass through each other; leading to the conclusion that the conjugacy classes of $\pi_1(\mathcal{M)}$ is the correct identifier of a defect. As such, we then calculated the structure of the manifold showing that $\pi_1\rm(SO(3)/A_4)\cong T'$, thus deducing that the conjugacy classes of $\rm T'$ is what needs to be looked at. Analyzing the structure of $\rm T'$, we demonstrated that multiple conjugacy classes can correspond to the same string network, since strings and anti-strings is a matter of orientation and they must have the same tension. We then calculated the class product which provides insight into the ways in which strings can interact. 
\par We then reviewed the non-Abelian VOS equations, and pointed out what has been left as a free parameter $\tilde{d}^k_{ij}$ can actually be written as $\tilde{d}^k_{ij}=P^k_{ij}\tilde{d}$ where $P^k_{ik}$ is the probability that a string of types $i$ and $j$ will produce a string of type $k$, which is derived from $\pi_1(\mathcal{M})$, and $\tilde{d}$ is a free parameter that contains kinematic information of the network. We provided sample calculations to demonstrate how $P^k_{ij}$ is found from the group structure of our manifold. Interestingly, network 2, corresponding to the centralizer of $\rm T'$, behaves like an Abelian string in the absence of bridging, as it reconnects with itself and passes through everything else. It turns out that the gravitational wave spectrum is enhanced due to other networks bridging into network 2. 

\par 
Using the derived probability coefficients, we can then solve the VOS equations and determine the long term behavior of the string networks in the limit of bridging and no bridging. We find that for each network, a scaling solution is met and the correlation length is smaller than the Abelian network. This is because of the bridging effects that produce a higher number density of strings. From the VOS evolution, we are able to calculate the gravitational wave spectrum produced by loops of string. After reviewing how gravitational waves are produced, we show the resulting spectra. The dominant contributions are found to be networks 1 and 2 since network 2 reconnects with probability 1 and all other interactions bridge into network 1. So we end up with a higher population of network 1 and network 2 strings, and hence a higher population of loops, and hence a more powerful signal. 

\par 
For future work, it would be interesting to apply this to a specific potential so that we can calculate the tension from first principles and potentially derive bounds on the parameters of the potential. There are also connections to non-invertible symmetries which is an interesting avenue to pursue~\cite{Abe:2026grk,Abe:2026zrk}. It would also be interesting to do this in the context of an explicit neutrino mass model, or a general flavor model, thereby connecting high energy particle theory with astrophysical observations. 

\section*{Acknowledgments}
We are thankful for useful discussions with Andrew Evans and Harold Matias. CMS acknowledges the support of the U.S. National Science Foundation (NSF) Graduate Research Fellowship Program (GRFP) and the Eugene Cota-Robles Fellowship provided by the University of California, Irvine

\appendix
\section{Appendix A: Interactions of A$_4$ Cosmic Strings} 
We will now go through the explicit calculations of all other string interactions, so that the derivation of Table \ref{tab:2T_bridges} can be traced. The calculation will be for the following interactions: (1,3), (1,4), (3,3), (3,4), and (4,4).  
\par For (1,3) the representative element is $s$ with centralizer $\{e, s, r, rs\}$. Note that no element in network 3 is in the centralizer of $s$, and hence there will be a bridge that forms. We will choose a representative from network 3 in $C_4''$ to interact with a representative from network 1. Since network 3 is comprised of the union of classes $C_4''\cup C_4'''$ one might be concerned about only choosing a representative from $C_4''$. To put this at ease note the following if $a\in C_6$ and $b\in C_4''$ then to find out where the flux lies we have $[a,b]=aba^{-1}b^{-1}=(aba^{-1})b^{-1}$, where $(aba^{-1})\in C_4''$ and $b^{-1}\in C_4'''$ thus the commutator must land in the product of $C_4''\times C_4'''$. Now lets assume the other way, which $b\in C_4'''$, then $[a,b]=aba^{-1}b^{-1}=(aba^{-1})b^{-1}$, where $(aba^{-1})\in C_4'''$ and $b\in C_4''$ and the commutator still lands in $C_4''\times C_4'''=4C_1+2C_6$ and thus it doesn't matter to choose a representative in $C_4''$ versus $C_4'''$. 
\par Further, we could also have $a\in C_4''$ and $b\in C_6$ then we would have the flux landing in $[a,b]=(aba^{-1})b^{-1}$ where $aba^{-1}\in C_6''$ and $b^{-1} \in C_6$ and the flux would land in $C_6 \times C_6=6C_1 +6C_1'+4C_6$ which is vastly different than $C_4'' \times C_4'''$. So there is ambiguity as to whether the flux lands in $C_1'$ or $C_6$. To resolve this we can show that $[a,b]\neq r$. We first assume that $[a,b]=r$ then: 
\begin{align*}
    aba^{-1}b^{-1}&=r\\ (aba^{-1}b^{-1})^2=r^2&=e \\
    aba^{-1}b^{-1}aba^{-1}b^{-1}&=e
\end{align*} which is only possible if $a=b$ which is not true by assumption and if b is in the centralizer of a, which is also not true. Thus $[a,b]\neq r$, so the flux must land in $C_6$, and we find that $P_{13}^1=1$. The same logic will apply to (1,4). 

\par For the case of (3,3), we will choose a representative of $rt$. Since network 3 both $rt$ and $rt^2=(rt)^{-1}$, two of the eight fluxes in network 3 will reconnect and $P_{\rm rec}=\frac{1}{4}$. The rest of the interactions will create bridges, and we find that $C_4''\times C_4'''=4C_1+2C_6$ so all the branching goes to network 1 and therefore $P_{33}^1=\frac{3}{4}$. The same holds if we looked at $[a,b]$ where $a,b\in C_4''$ or $[a,b]\in C_4'''$.

\par For (3,4) we choose our representative to be $rt$ once again. We have $t,t^2$ from network 4 in the centralizer of $rt$ and therefore two of the eight fluxes will pass through, so the rest must form a bridge. Since $C_4''\times C_4=4C_1+2C_6$ we have $P_{34}^1=\frac{3}{4}$.
\par For the case of (4,4) since $t=t$ and $t^{-1}=t^2$, we have two of eight fluxes reconnecting $P_{\rm reco}=\frac{1}{4}$. Since the bridge resulting from the flux of two $C_4$ strings must lie in $C_4 \times C_4'=4C_1+2C_6$, we have $P_{34}^1=\frac{3}{4}$.
\label{appendixA}

\section{Appendix B: Evolution and GW of String Networks With Bridging}
\label{appendixB}
In this appendix we present the correlation length, energy density, and gravitational wave spectrum for networks 3 and 4 which were omitted from the main text. Since networks 3 and 4 have the exact same bridging and reconnecting probabilities, they have identical curves and therefore we will only show plots from network 3.  
\begin{figure}[h]
    \centering
    \includegraphics[width=\textwidth]{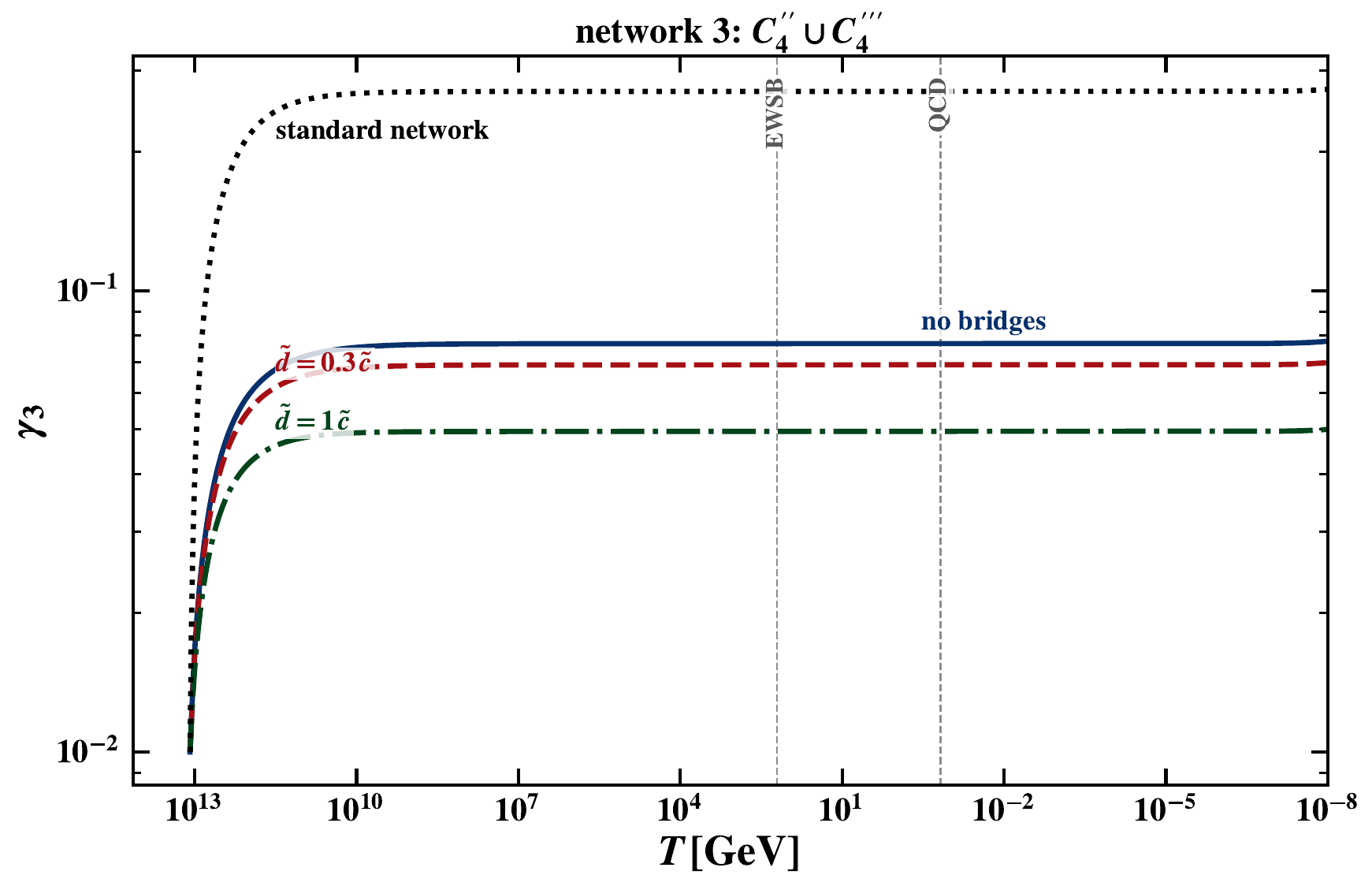}
    \caption{Dimensionless correlation length evolution for network 3.}
    \label{fig:gw_net3}
\end{figure}
\begin{figure}[h]
    \centering
    \includegraphics[width=\textwidth]{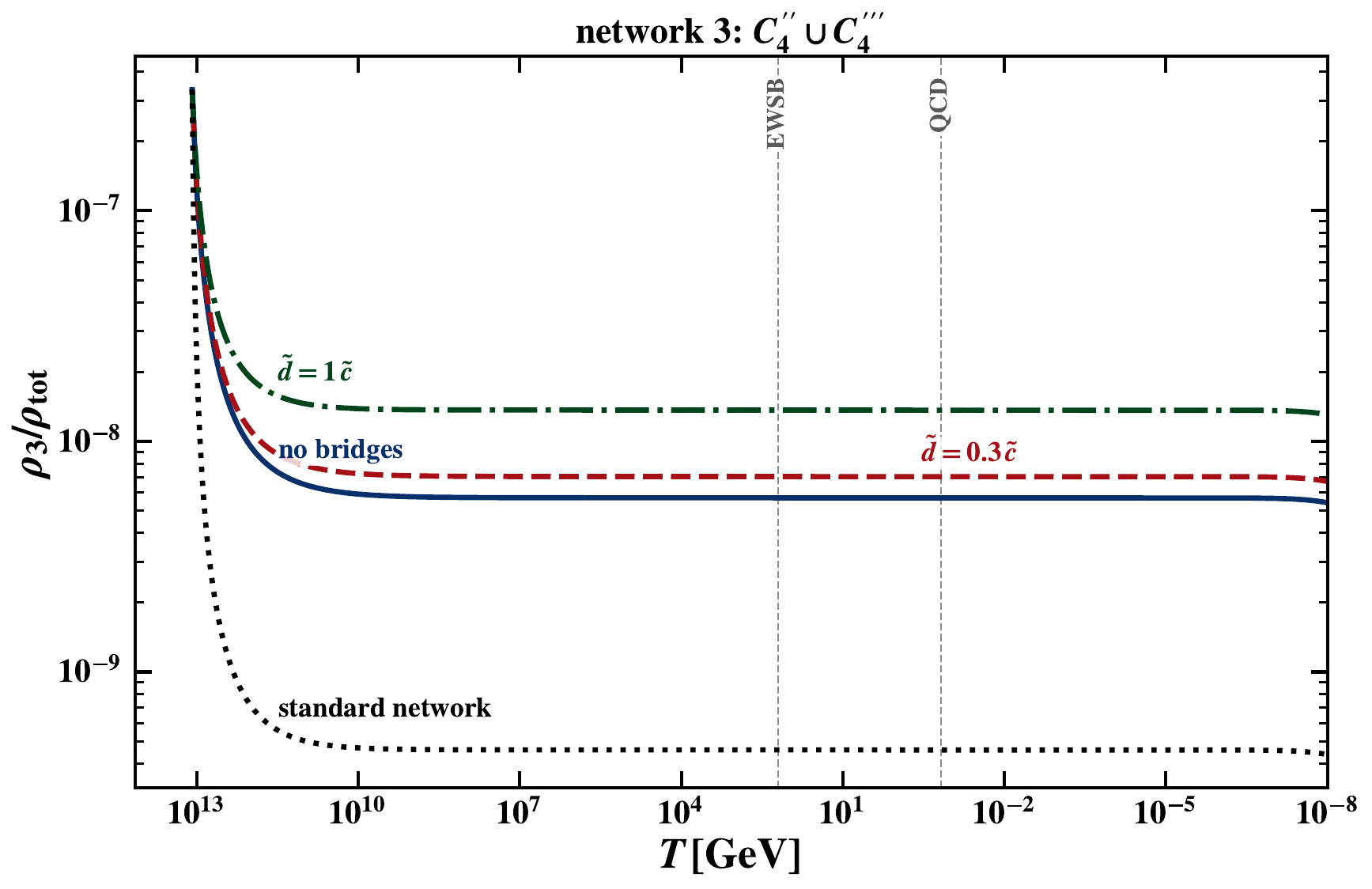}
     \caption{Energy density for network 3.}
    \label{fig:rho_net3}
\end{figure}
\begin{figure}[h]
    \centering
    \includegraphics[width=\textwidth]{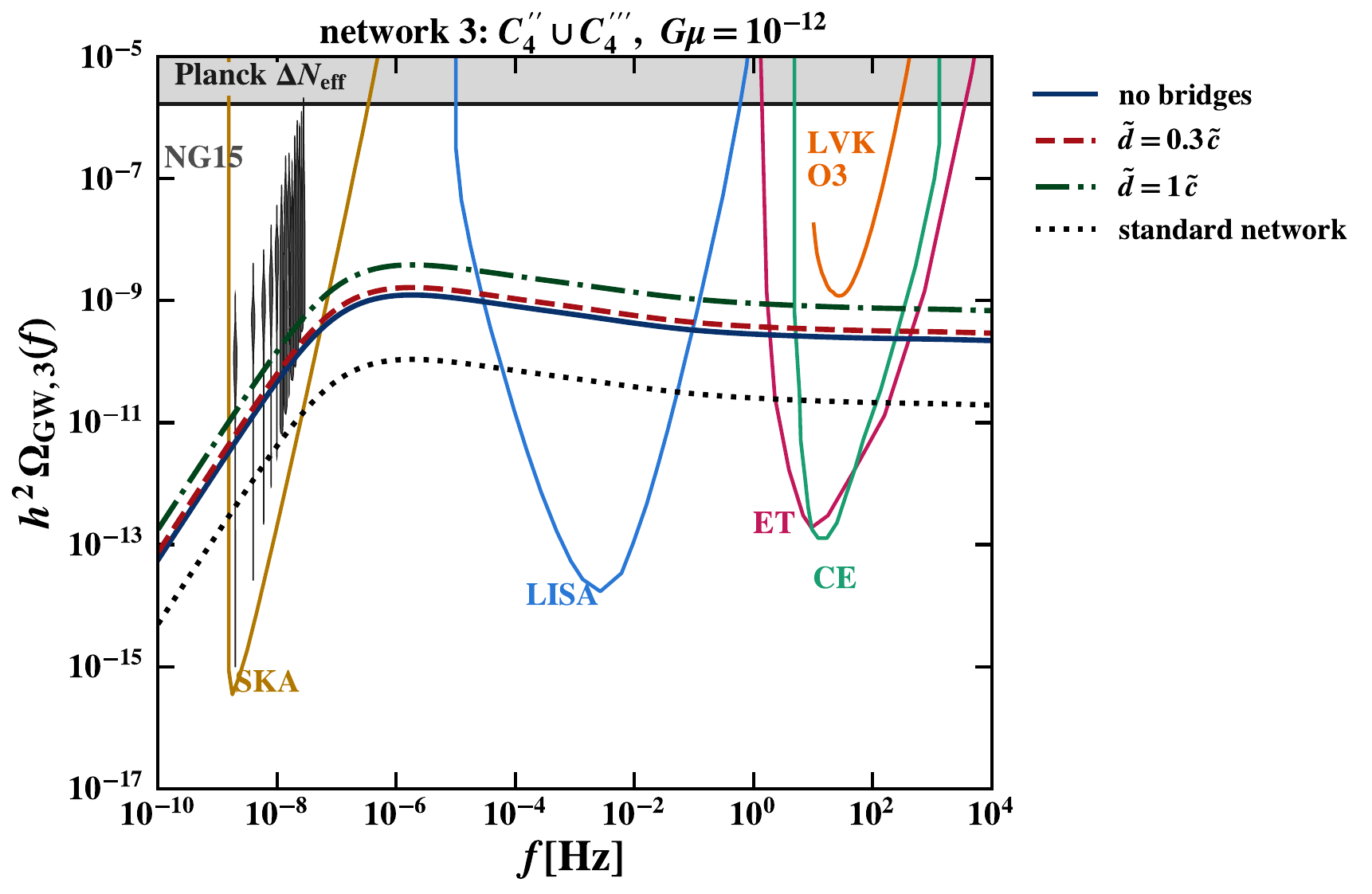}
     \caption{GW spectrum from network 3.}
    \label{fig:gw_net4}
\end{figure}
\FloatBarrier
\bibliographystyle{JHEP}     
\bibliography{References}

@article{Notari:2025kqq,
    author = "Notari, Alessio and Rompineve, Fabrizio and Torrenti, Francisco",
    title = "{The spectrum of gravitational waves from annihilating domain walls}",
    eprint = "2504.03636",
    archivePrefix = "arXiv",
    primaryClass = "astro-ph.CO",
    doi = "10.1088/1475-7516/2025/07/049",
    journal = "JCAP",
    volume = "07",
    pages = "049",
    year = "2025"
}

@article{Fu:2024jhu,
    author = "Fu, Bowen and King, Stephen F. and Marsili, Luca and Pascoli, Silvia and Turner, Jessica and Zhou, Ye-Ling",
    title = "{Non-Abelian domain walls and gravitational waves}",
    eprint = "2409.16359",
    archivePrefix = "arXiv",
    primaryClass = "hep-ph",
    reportNumber = "IPPP/24/62",
    doi = "10.1007/JHEP04(2025)142",
    journal = "JHEP",
    volume = "04",
    pages = "142",
    year = "2025"
}

@article{Saikawa:2017hiv,
    author = "Saikawa, Ken'ichi",
    title = "{A review of gravitational waves from cosmic domain walls}",
    eprint = "1703.02576",
    archivePrefix = "arXiv",
    primaryClass = "hep-ph",
    reportNumber = "DESY-17-036",
    doi = "10.3390/universe3020040",
    journal = "Universe",
    volume = "3",
    number = "2",
    pages = "40",
    year = "2017"
}

@article{Gelmini:2020bqg,
    author = "Gelmini, Graciela B. and Pascoli, Silvia and Vitagliano, Edoardo and Zhou, Ye-Ling",
    title = "{Gravitational wave signatures from discrete flavor symmetries}",
    eprint = "2009.01903",
    archivePrefix = "arXiv",
    primaryClass = "hep-ph",
    doi = "10.1088/1475-7516/2021/02/032",
    journal = "JCAP",
    volume = "02",
    pages = "032",
    year = "2021"
}

@article{Jueid:2023cgp,
    author = "Jueid, Adil and Loualidi, Mohamed Amin and Nasri, Salah and Ouahid, Mohamed Amine",
    title = "{Cosmological domain walls from the breaking of S4 flavor symmetry}",
    eprint = "2312.04388",
    archivePrefix = "arXiv",
    primaryClass = "hep-ph",
    reportNumber = "CTPU-PTC-23-52",
    doi = "10.1103/PhysRevD.109.055048",
    journal = "Phys. Rev. D",
    volume = "109",
    number = "5",
    pages = "055048",
    year = "2024"
}

@article{Ma:2001dn,
    author = "Ma, Ernest and Rajasekaran, G.",
    title = "{Softly broken A(4) symmetry for nearly degenerate neutrino masses}",
    eprint = "hep-ph/0106291",
    archivePrefix = "arXiv",
    reportNumber = "UCRHEP-T308",
    doi = "10.1103/PhysRevD.64.113012",
    journal = "Phys. Rev. D",
    volume = "64",
    pages = "113012",
    year = "2001"
}

@article{Chigusa:2018hhl,
    author = "Chigusa, So and Nakayama, Kazunori",
    title = "{Anomalous Discrete Flavor Symmetry and Domain Wall Problem}",
    eprint = "1808.09601",
    archivePrefix = "arXiv",
    primaryClass = "hep-ph",
    reportNumber = "UT-18-18",
    doi = "10.1016/j.physletb.2018.11.027",
    journal = "Phys. Lett. B",
    volume = "788",
    pages = "249--255",
    year = "2019"
}

@article{Preskill:1991kd,
    author = "Preskill, John and Trivedi, Sandip P. and Wilczek, Frank and Wise, Mark B.",
    title = "{Cosmology and broken discrete symmetry}",
    reportNumber = "IASSNS-HEP-91-11, CALT-68-1718",
    doi = "10.1016/0550-3213(91)90241-O",
    journal = "Nucl. Phys. B",
    volume = "363",
    pages = "207--220",
    year = "1991"
}

@article{Zeldovich:1974uw,
    author = "Zeldovich, Ya. B. and Kobzarev, I. Yu. and Okun, L. B.",
    title = "{Cosmological Consequences of the Spontaneous Breakdown of Discrete Symmetry}",
    reportNumber = "SLAC-TRANS-0165, IPM-MOSCOW-15",
    journal = "Zh. Eksp. Teor. Fiz.",
    volume = "67",
    pages = "3--11",
    year = "1974"
}

@article{Vilenkin:1981zs,
    author = "Vilenkin, A.",
    title = "{Gravitational Field of Vacuum Domain Walls and Strings}",
    doi = "10.1103/PhysRevD.23.852",
    journal = "Phys. Rev. D",
    volume = "23",
    pages = "852--857",
    year = "1981"
}

@article{Ghoshal:2025gci,
    author = "Ghoshal, Anish and Hamada, Yu",
    title = "{Gravitational waves created by current-carrying domain walls}",
    eprint = "2501.01542",
    archivePrefix = "arXiv",
    primaryClass = "hep-ph",
    reportNumber = "DESY-24-198",
    doi = "10.1103/24w5-rflt",
    journal = "Phys. Rev. D",
    volume = "112",
    number = "11",
    pages = "115019",
    year = "2025"
}

@article{Feruglio:2008ht,
    author = "Feruglio, Ferruccio and Hagedorn, Claudia and Lin, Yin and Merlo, Luca",
    title = "{Lepton Flavour Violation in Models with A(4) Flavour Symmetry}",
    eprint = "0807.3160",
    archivePrefix = "arXiv",
    primaryClass = "hep-ph",
    reportNumber = "DFPD-08-TH-09",
    doi = "10.1016/j.nuclphysb.2008.10.002",
    journal = "Nucl. Phys. B",
    volume = "809",
    pages = "218--243",
    year = "2009"
}

@article{Feruglio:2017spp,
    author = "Feruglio, Ferruccio",
    title = "Are neutrino masses modular forms?",
    eprint = "1706.08749",
    archivePrefix = "arXiv",
    primaryClass = "hep-ph",
    year = "2019",
    journal = "In *From my vast repertoire ... Guido Altarelli's legacy*",
    pages = "227--266",
    doi = "10.1142/9789813238053_0012"
}

@article{Froggatt:1978nt,
    author = "Froggatt, C. D. and Nielsen, Holger Bech",
    title = "{Hierarchy of Quark Masses, Cabibbo Angles and CP Violation}",
    reportNumber = "CERN-TH-2519",
    doi = "10.1016/0550-3213(79)90316-X",
    journal = "Nucl. Phys. B",
    volume = "147",
    pages = "277--298",
    year = "1979"
}

@article{Almumin:2022rml,
    author = "Almumin, Yahya and Chen, Mu-Chun and Cheng, Murong and Knapp-Perez, Victor and Li, Yulun and Mondol, Adreja and Ramos-Sanchez, Saul and Ratz, Michael and Shukla, Shreya",
    title = "{Neutrino Flavor Model Building and the Origins of Flavor and CP Violation}",
    eprint = "2204.08668",
    archivePrefix = "arXiv",
    primaryClass = "hep-ph",
    reportNumber = "UCI-TR-2022-03",
    doi = "10.3390/universe9120512",
    journal = "Universe",
    volume = "9",
    number = "12",
    pages = "512",
    year = "2023"
}

@article{Kobayashi:2023zzc,
    author = "Kobayashi, Tatsuo and Tanimoto, Morimitsu",
    editor = {Buchalla, Gerhard and L{\"u}st, Dieter and Xing, Zhi-zhong},
    title = "{Modular flavor symmetric models}",
    eprint = "2307.03384",
    archivePrefix = "arXiv",
    primaryClass = "hep-ph",
    doi = "10.1142/S0217751X24410124",
    journal = "Int. J. Mod. Phys. A",
    volume = "39",
    number = "09n10",
    pages = "2441012",
    year = "2024"
}

@misc{Ghoshal:2026hev,
    author = "Ghoshal, Anish and Pal, Pratyay",
    title = "{Cosmic Collider Gravitational Waves sourced by Right-handed Neutrino production from Bubbles: Testing Seesaw, Leptogenesis and Dark Matter}",
    eprint = "2601.02458",
    archivePrefix = "arXiv",
    primaryClass = "astro-ph.CO",
    month = "1",
    year = "2026",
    note = "arXiv:2601.02458 [astro-ph.CO]"
}

@article{Dror:2019syi,
    author = "Dror, Jeff A. and Hiramatsu, Takashi and Kohri, Kazunori and Murayama, Hitoshi and White, Graham",
    title = "{Testing the Seesaw Mechanism and Leptogenesis with Gravitational Waves}",
    eprint = "1908.03227",
    archivePrefix = "arXiv",
    primaryClass = "hep-ph",
    reportNumber = "IPMU19-0108, DESY-19-138, DESY 19-138, KEK-TH-2147, KEK-Cosmo-241",
    doi = "10.1103/PhysRevLett.124.041804",
    journal = "Phys. Rev. Lett.",
    volume = "124",
    number = "4",
    pages = "041804",
    year = "2020"
}

@article{Fu:2025qhf,
    author = "Fu, Bowen and King, Stephen F. and Marsili, Luca and Turner, Jessica and Zhou, Ye-Ling",
    title = "{Domain Walls in $A_4$ Flavour Models}",
    eprint = "2512.13784",
    archivePrefix = "arXiv",
    primaryClass = "hep-ph",
    reportNumber = "IPPP/25/80",
    month = "12",
    year = "2025",
    note = "arXiv:2512.13784 [hep-ph]"
}

@article{Barman:2022yos,
    author = "Barman, Basabendu and Borah, Debasish and Dasgupta, Arnab and Ghoshal, Anish",
    title = "{Probing high scale Dirac leptogenesis via gravitational waves from domain walls}",
    eprint = "2205.03422",
    archivePrefix = "arXiv",
    primaryClass = "hep-ph",
    doi = "10.1103/PhysRevD.106.015007",
    journal = "Phys. Rev. D",
    volume = "106",
    number = "1",
    pages = "015007",
    year = "2022"
}

@article{Jana:2025vyb,
    author = "Jana, Sudip and Manna, Sudip and K, Vishnu P.",
    title = "{Gravitational Wave Signature and the Nature of Neutrino Masses: Majorana, Dirac, or Pseudo-Dirac?}",
    eprint = "2509.10456",
    archivePrefix = "arXiv",
    primaryClass = "hep-ph",
    reportNumber = "HRI-RECAPP-2025-09, MS-TP-25-19",
    month = "9",
    year = "2025"
}

@article{Paul:2020wbz,
    author = "Paul, Avik and Mukhopadhyay, Upala and Majumdar, Debasish",
    title = "{Gravitational Wave Signatures from Domain Wall and Strong First-Order Phase Transitions in a Two Complex Scalar extension of the Standard Model}",
    eprint = "2010.03439",
    archivePrefix = "arXiv",
    primaryClass = "hep-ph",
    doi = "10.1007/JHEP05(2021)223",
    journal = "JHEP",
    volume = "05",
    pages = "223",
    year = "2021"
}

@article{King:2018fke,
    author = "King, Stephen F. and Zhou, Ye-Ling",
    title = "{Spontaneous breaking of $SO(3)$ to finite family symmetries with supersymmetry - an $A_4$ model}",
    eprint = "1809.10292",
    archivePrefix = "arXiv",
    primaryClass = "hep-ph",
    reportNumber = "IPPP/18/83",
    doi = "10.1007/JHEP11(2018)173",
    journal = "JHEP",
    volume = "11",
    pages = "173",
    year = "2018"
}

@book{Vilenkin:2000jqa,
    author = "Vilenkin, A. and Shellard, E. P. S.",
    title = "{Cosmic Strings and Other Topological Defects}",
    isbn = "978-0-521-65476-0",
    publisher = "Cambridge University Press",
    month = "7",
    year = "2000"
}

@article{Dunsky:2021tih,
    author = "Dunsky, David I. and Ghoshal, Anish and Murayama, Hitoshi and Sakakihara, Yuki and White, Graham",
    title = "{GUTs, hybrid topological defects, and gravitational waves}",
    eprint = "2111.08750",
    archivePrefix = "arXiv",
    primaryClass = "hep-ph",
    doi = "10.1103/PhysRevD.106.075030",
    journal = "Phys. Rev. D",
    volume = "106",
    number = "7",
    pages = "075030",
    year = "2022"
}

@article{Antusch:2026msw,
    author = "Antusch, Stefan and de Medeiros Varzielas, Ivo and Levy, Miguel",
    title = "{Gravity tidings from domain walls: Flavour hierarchies are making waves}",
    eprint = "2603.23395",
    archivePrefix = "arXiv",
    primaryClass = "hep-ph",
    month = "3",
    year = "2026"
}

@article{Barreto:2026igt,
    author = "Barreto, Gon{\c{c}}alo and de Medeiros Varzielas, Ivo and Zhou, Ye-Ling",
    title = "{Domain Walls from $\Sigma(36 \times 3)$, $\Delta(54)$ and $\Delta(27)$ potentials}",
    eprint = "2603.04496",
    archivePrefix = "arXiv",
    primaryClass = "hep-ph",
    month = "3",
    year = "2026"
}

@article{McGraw:1996py,
    author = "McGraw, Patrick",
    title = "{Dynamical simulation of nonAbelian cosmic strings}",
    eprint = "hep-th/9603153",
    archivePrefix = "arXiv",
    reportNumber = "CALT-68-2044",
    month = "3",
    year = "1996"
}

@article{McGraw:1997nx,
    author = "McGraw, Patrick",
    title = "{Evolution of a nonAbelian cosmic string network}",
    eprint = "astro-ph/9706182",
    archivePrefix = "arXiv",
    doi = "10.1103/PhysRevD.57.3317",
    journal = "Phys. Rev. D",
    volume = "57",
    pages = "3317--3339",
    year = "1998"
}

@article{Alford:1992yx,
    author = "Alford, Mark G. and Lee, Kai-Ming and March-Russell, John and Preskill, John",
    title = "{Quantum field theory of nonAbelian strings and vortices}",
    eprint = "hep-th/9112038",
    archivePrefix = "arXiv",
    reportNumber = "PUPT-91-1288, CALT-68-1700, NSF-ITP-91-128",
    doi = "10.1016/0550-3213(92)90468-Q",
    journal = "Nucl. Phys. B",
    volume = "384",
    pages = "251--317",
    year = "1992"
}

@article{Ishimori:2010au,
    author = "Ishimori, Hajime and Kobayashi, Tatsuo and Ohki, Hiroshi and Shimizu, Yusuke and Okada, Hiroshi and Tanimoto, Morimitsu",
    title = "{Non-Abelian Discrete Symmetries in Particle Physics}",
    eprint = "1003.3552",
    archivePrefix = "arXiv",
    primaryClass = "hep-th",
    reportNumber = "KUNS-2260",
    doi = "10.1143/PTPS.183.1",
    journal = "Prog. Theor. Phys. Suppl.",
    volume = "183",
    pages = "1--163",
    year = "2010"
}

@article{Avgoustidis:2007aa,
    author = "Avgoustidis, A. and Shellard, E. P. S.",
    title = "{Velocity-Dependent Models for Non-Abelian/Entangled String Networks}",
    eprint = "0705.3395",
    archivePrefix = "arXiv",
    primaryClass = "astro-ph",
    reportNumber = "UB-ECM-PF-07-11, DAMTP-2007-46",
    doi = "10.1103/PhysRevD.78.103510",
    journal = "Phys. Rev. D",
    volume = "78",
    pages = "103510",
    year = "2008",
    note = "[Erratum: Phys.Rev.D 80, 129907 (2009)]"
}

@article{Abe:2026zrk,
    author = "Abe, Yoshihiko and Higaki, Tetsutaro and Murakami, Kazuya and Nitta, Muneto and Yokokura, Ryo",
    title = "{Non-Abelian $A_4$ vortices in $SO(3)$ gauge theory and non-invertible symmetries}",
    eprint = "2607.12366",
    archivePrefix = "arXiv",
    primaryClass = "hep-th",
    month = "7",
    year = "2026"
}

@manual{GAP4,
    organization = "The GAP~Group",
    title        = "{GAP -- Groups, Algorithms, and Programming,
                    Version 4.16.1}",
    year         = 2026,
    url          = "\url{https://www.gap-system.org}",
    }

@article{Nakamura:2025ptd,
    author = "Nakamura, Kazuto and Yamada, Masaki",
    title = "{Delayed scaling of multitype cosmic $F$- and $D$-strings in velocity-dependent one-scale models}",
    eprint = "2512.16821",
    archivePrefix = "arXiv",
    primaryClass = "hep-ph",
    reportNumber = "TU-1293",
    doi = "10.1103/4tkt-pfk2",
    journal = "Phys. Rev. D",
    volume = "113",
    number = "6",
    pages = "063537",
    year = "2026"
}

@article{Martins:2004vs,
    author = "Martins, C. J. A. P.",
    title = "{Scaling laws for non-intercommuting cosmic string networks}",
    eprint = "hep-ph/0410326",
    archivePrefix = "arXiv",
    reportNumber = "DAMTP-2004-117",
    doi = "10.1103/PhysRevD.70.107302",
    journal = "Phys. Rev. D",
    volume = "70",
    pages = "107302",
    year = "2004"
}

@article{Tye:2005fn,
    author = "Tye, S. -H. Henry and Wasserman, Ira and Wyman, Mark",
    title = "{Scaling of multi-tension cosmic superstring networks}",
    eprint = "astro-ph/0503506",
    archivePrefix = "arXiv",
    doi = "10.1103/PhysRevD.71.103508",
    journal = "Phys. Rev. D",
    volume = "71",
    pages = "103508",
    year = "2005",
    note = "[Erratum: Phys.Rev.D 71, 129906 (2005)]"
}

@article{Copeland:2005cy,
    author = "Copeland, Edmund J. and Saffin, P. M.",
    title = "{On the evolution of cosmic-superstring networks}",
    eprint = "hep-th/0505110",
    archivePrefix = "arXiv",
    doi = "10.1088/1126-6708/2005/11/023",
    journal = "JHEP",
    volume = "11",
    pages = "023",
    year = "2005"
}

@article{Copeland:2006eh,
    author = "Copeland, E. J. and Kibble, T. W. B. and Steer, Daniele A.",
    title = "{Collisions of strings with Y junctions}",
    eprint = "hep-th/0601153",
    archivePrefix = "arXiv",
    doi = "10.1103/PhysRevLett.97.021602",
    journal = "Phys. Rev. Lett.",
    volume = "97",
    pages = "021602",
    year = "2006"
}

@article{Saffin:2005cs,
    author = "Saffin, P. M.",
    title = "{A Practical model for cosmic (p,q) superstrings}",
    eprint = "hep-th/0506138",
    archivePrefix = "arXiv",
    doi = "10.1088/1126-6708/2005/09/011",
    journal = "JHEP",
    volume = "09",
    pages = "011",
    year = "2005"
}

@article{Hindmarsh:2006qn,
    author = "Hindmarsh, Mark and Saffin, P. M.",
    title = "{Scaling in a SU(2)$/\mathbb{Z}_{3}$ model of cosmic superstring networks}",
    eprint = "hep-th/0605014",
    archivePrefix = "arXiv",
    doi = "10.1088/1126-6708/2006/08/066",
    journal = "JHEP",
    volume = "08",
    pages = "066",
    year = "2006"
}

@article{Martins:1995tg,
    author = "Martins, C. J. A. P. and Shellard, E. P. S.",
    title = "{String evolution with friction}",
    eprint = "hep-ph/9507335",
    archivePrefix = "arXiv",
    reportNumber = "DAMTP-R-95-10",
    doi = "10.1103/PhysRevD.53.R575",
    journal = "Phys. Rev. D",
    volume = "53",
    pages = "575--579",
    year = "1996"
}

@article{Martins:2000cs,
    author = "Martins, C. J. A. P. and Shellard, E. P. S.",
    title = "{Extending the velocity dependent one scale string evolution model}",
    eprint = "hep-ph/0003298",
    archivePrefix = "arXiv",
    doi = "10.1103/PhysRevD.65.043514",
    journal = "Phys. Rev. D",
    volume = "65",
    pages = "043514",
    year = "2002"
}

@article{Gouttenoire:2019kij,
    author = "Gouttenoire, Yann and Servant, G{\'e}raldine and Simakachorn, Peera",
    title = "{Beyond the Standard Models with Cosmic Strings}",
    eprint = "1912.02569",
    archivePrefix = "arXiv",
    primaryClass = "hep-ph",
    reportNumber = "DESY-19-204",
    doi = "10.1088/1475-7516/2020/07/032",
    journal = "JCAP",
    volume = "07",
    pages = "032",
    year = "2020"
}

@article{Blanco-Pillado:2013qja,
    author = "Blanco-Pillado, Jose J. and Olum, Ken D. and Shlaer, Benjamin",
    title = "{The number of cosmic string loops}",
    eprint = "1309.6637",
    archivePrefix = "arXiv",
    primaryClass = "astro-ph.CO",
    doi = "10.1103/PhysRevD.89.023512",
    journal = "Phys. Rev. D",
    volume = "89",
    number = "2",
    pages = "023512",
    year = "2014"
}

@article{Blanco-Pillado:2017rnf,
    author = "Blanco-Pillado, Jose J. and Olum, Ken D. and Siemens, Xavier",
    title = "{New limits on cosmic strings from gravitational wave observation}",
    eprint = "1709.02434",
    archivePrefix = "arXiv",
    primaryClass = "astro-ph.CO",
    doi = "10.1016/j.physletb.2018.01.050",
    journal = "Phys. Lett. B",
    volume = "778",
    pages = "392--396",
    year = "2018"
}

@article{Cella:2015btn,
    author = "Cella, G. and Pieroni, M.",
    title = "{A simple model for the evolution of a non-Abelian cosmic string network}",
    eprint = "1512.02117",
    archivePrefix = "arXiv",
    primaryClass = "astro-ph.CO",
    doi = "10.1088/1475-7516/2016/06/034",
    journal = "JCAP",
    volume = "06",
    pages = "034",
    year = "2016"
}

@article{Buchmuller:2019gfy,
    author = "Buchmuller, Wilfried and Domcke, Valerie and Murayama, Hitoshi and Schmitz, Kai",
    title = "{Probing the scale of grand unification with gravitational waves}",
    eprint = "1912.03695",
    archivePrefix = "arXiv",
    primaryClass = "hep-ph",
    reportNumber = "CERN-TH-2019-215, DESY-19-210, DESY 19-210, IPMU 19-0179",
    doi = "10.1016/j.physletb.2020.135764",
    journal = "Phys. Lett. B",
    volume = "809",
    pages = "135764",
    year = "2020"
}

@article{Priya:2026vpo,
    author = "Priya and Singh, Labh and Chauhan, B. C. and Verma, Surender",
    title = "{Radiative Lifting of $\mathbb{Z}_3$ Domain-Wall Degeneracy in a Type-III Seesaw Model: Implications for Leptogenesis and Gravitational Waves}",
    eprint = "2606.22642",
    archivePrefix = "arXiv",
    primaryClass = "hep-ph",
    month = "6",
    year = "2026"
}

@article{Hua:2026mgn,
    author = "Hua, Jianjun and Fu, Bowen and Tang, Yi-Lei",
    title = "{One-Dimensional Simulations of the Topological Defects in a 3:1 $U(1)$ Model}",
    eprint = "2607.13066",
    archivePrefix = "arXiv",
    primaryClass = "hep-ph",
    month = "7",
    year = "2026"
}

@article{Puyam:2026eyi,
    author = "Puyam, Victoria and Das, Mrinal Kumar",
    title = "{Neutrino Mass and its Impact on Gravitational Waves from Domain Wall Collision}",
    eprint = "2608.08084",
    archivePrefix = "arXiv",
    primaryClass = "hep-ph",
    month = "8",
    year = "2026"
}

@article{Abe:2026grk,
    author = "Abe, Yoshihiko and Higaki, Tetsutaro and Murakami, Kazuya and Nitta, Muneto and Yokokura, Ryo",
    title = "{The stability of non-Abelian $A_4$ vortices in $SO(3)$ gauge theory}",
    eprint = "2610.19963",
    archivePrefix = "arXiv",
    primaryClass = "hep-th",
    month = "10",
    year = "2026"
}
\end{document}